\documentclass[prx,aps,twocolumn,superscriptaddress]{revtex4-2}

\usepackage[pdftex]{graphicx}
\usepackage{amsbsy,amssymb,amsmath,bm,mathtools}
\usepackage{amsfonts}

\usepackage{xspace}
\usepackage{bm}
\usepackage{color}
 
\usepackage{url}
\usepackage[section]{placeins}
\usepackage[colorlinks=true,linkcolor=blue,citecolor=blue]{hyperref}

\usepackage[mathscr]{euscript}

\usepackage{graphicx}
\usepackage{dcolumn}

\begin{document}

\preprint{APS/123-QED}

\title{Machine-learning modeling of magnetization dynamics in quasi-equilibrium and driven metallic spin systems}

\author{Gia-Wei Chern}
\affiliation{Department of Physics, University of Virginia, Charlottesville, VA 22904, USA}

\author{Yunhao Fan}
\affiliation{Department of Physics, University of Virginia, Charlottesville, VA 22904, USA}

\author{Sheng Zhang}
\affiliation{Department of Physics, University of Virginia, Charlottesville, VA 22904, USA}

\author{Puhan Zhang}
\affiliation{Department of Physics, University of Virginia, Charlottesville, VA 22904, USA}

\begin{abstract}
We review recent advances in machine-learning (ML) force-field methods for large-scale Landau-Lifshitz-Gilbert (LLG) simulations of metallic spin systems. We generalize the Behler-Parrinello (BP) ML architecture---originally developed for quantum molecular dynamics---to construct scalable and transferable ML models capable of capturing the intricate dependence of electron-mediated exchange fields on the local magnetic environment characteristic of itinerant magnets. A central ingredient of this framework is the implementation of symmetry-aware magnetic descriptors based on group-theoretical bispectrum formalisms. Leveraging these ML force fields, LLG simulations faithfully reproduce hallmark non-collinear magnetic orders---such as the $120^\circ$ and tetrahedral states---on the triangular lattice, and successfully capture the complex spin textures emerging in the mixed-phase states of a square-lattice double-exchange model under thermal quench. We further discuss a generalized potential theory that extends the BP formalism to incorporate both conservative and nonconservative electronic torques, thereby enabling ML models to learn nonequilibrium exchange fields from computationally demanding microscopic approaches such as nonequilibrium Green’s-function techniques. This extension yields quantitatively accurate predictions of voltage-driven domain-wall motion and establishes a foundation for quantum-accurate, multiscale modeling of nonequilibrium spin dynamics and spintronic functionalities.
\end{abstract}

\maketitle


\section{Introduction}

\label{sec:intro}

Over the past two decades, machine learning (ML) has transformed a wide range of disciplines, spanning industrial applications, data-driven engineering, and fundamental scientific research. In physics and materials sciences, the integration of ML techniques has not only yielded numerous remarkable successes but has also opened entirely new paradigms for exploring complex physical phenomena and accelerating theoretical discovery~\cite{carleo19,sarma19,morgan20,bedolla21,karagiorgi22,karniadakis21,carrasquilla17,carleo17,ramprasad17,gubernatis18,butler18,schmidt19,batra21,boehnlein22,wang20}. The transformative power of ML arises from its ability to act as a universal approximator for high-dimensional, nonlinear mappings. This capability allows ML models to capture intricate structure–property relationships, learn effective representations directly from data, and generalize to previously unseen regimes. Consequently, ML has greatly enhanced the efficiency, scalability, and predictive accuracy of complex numerical simulations, enabling new levels of realism and precision in modeling physical systems.

Among these developments, one of the most prominent milestones has been the application of ML to {\em ab~initio} or quantum molecular dynamics (QMD)---in particular, the accurate prediction of atomic energies and forces~\cite{behler07,bartok10,li15,shapeev16,behler16,botu17,smith17,zhang18,deringer19,mcgibbon17,suwa19,chmiela17,chmiela18,sauceda20,unke21}. Unlike classical molecular dynamics based on empirical force fields, QMD computes atomic forces by explicitly integrating out the electronic degrees of freedom as the atomic configurations evolve~\cite{Marx09}. While this first-principles approach---often relying on many-body electronic structure methods such as density functional theory (DFT)---yields high fidelity, its substantial computational cost severely limits the accessible temporal and spatial scales. Machine-learning force-field models provide a powerful solution to this bottleneck by emulating the underlying electronic-structure calculations with near-quantum accuracy at a fraction of the cost. This synergy between ML and QMD has opened the door to large-scale, high-accuracy atomistic simulations, effectively bridging the gap between {\em ab initio} precision and mesoscale modeling.

Here, we review recent advances in scalable ML frameworks for simulating adiabatic magnetization dynamics in metallic spin systems, exemplified by the s-d exchange, double-exchange, and Kondo-lattice models~\cite{zhang20,zhang21,zhang23,cheng23b,Fan24,tyberg25}. In such itinerant magnets, spin dynamics are primarily governed by exchange interactions mediated by conduction electrons, leading to computational demands comparable to those of QMD. To evaluate the local effective magnetic fields that drive spin evolution, one must, in principle, solve the real-space electronic Hamiltonian corresponding to each instantaneous spin configuration at every integration step---a task that becomes prohibitively expensive for large systems. The ML force-field paradigm offers an efficient alternative by learning the mapping between spin configurations and their corresponding electronic energies or torques, enabling linear-scaling simulations with near–quantum accuracy at a computational cost comparable to classical spin dynamics.

Metallic spin systems are of great interest owing to their rich variety of coupled spin–charge textures stabilized by competing exchange interactions and electron-mediated correlations. These materials host emergent mesoscale structures—such as magnetic vortices, skyrmions, and stripe-like modulations—arising from the delicate balance between localized spins and itinerant electrons. Among these, magnetic vortices and skyrmions have attracted particular attention for their topological stability, particle-like dynamics, and potential applications in next-generation spintronic and neuromorphic devices~\cite{bogdanov89,rossler06,muhlbaure09,yu10,yu11,seki12,nagaosa13}. The associated Berry-phase effects act as emergent magnetic fields on conduction electrons, giving rise to unconventional transport responses such as the topological and anomalous Hall effects.

A striking example of electronically driven self-organization occurs in colossal magnetoresistance (CMR) manganites, where ferromagnetic metallic and antiferromagnetic or charge-ordered insulating domains coexist within a nanoscale mosaic~\cite{dagotto_book,dagotto05,moreo99,dagotto01,mathur03,nagaev02,fath99,renner02,salamon01}. These inhomogeneous textures are highly susceptible to external perturbations—magnetic field, pressure, or carrier doping—leading to dramatic, often nonlinear, changes in transport and magnetization. This intricate interplay among spin, charge, and lattice degrees of freedom exemplifies the emergent phenomena characteristic of strongly correlated itinerant magnets.

Taken together, these developments underscore the growing need for computational frameworks that can capture the full complexity of itinerant magnetism while remaining scalable to experimentally relevant length and time scales. ML force fields---building on the successes of their atomic counterparts---offer a powerful route toward this goal, enabling quantum-accurate simulations of magnetic textures, phase competition, and nonequilibrium spin dynamics at unprecedented scales. In this review, we survey the emerging methodologies, underlying theoretical principles, and representative applications of these ML-driven approaches, with the aim of providing both a unifying perspective and a roadmap for future advances at the interface of spin dynamics, electronic structure, and machine learning.

The remainder of this article is organized as follows. Section~\ref{sec:BP} introduces the ML force-field framework based on the Behler-Parrinello (BP) architecture for modeling spin dynamics in quasi-equilibrium systems. Section~\ref{sec:descriptor} presents the construction of symmetry-aware magnetic descriptors—an essential component of the framework---designed to preserve the symmetries of the underlying electronic Hamiltonian. Applications of the ML approach to representative noncollinear and noncoplanar magnetic orders are discussed in Sec.~\ref{sec:m-orders}. A dynamical example is provided in Sec.~\ref{sec:relaxation}, where the ML framework is used to model the thermal-quench evolution of mixed-phase states comprising coexisting ferromagnetic clusters and a N'eel-ordered background. Section~\ref{sec:generalized-BP} introduces the extension of the BP architecture to include nonconservative forces, relevant for driven nonequilibrium systems, and Sec.~\ref{sec:noneq} demonstrates this extended formalism in the context of a voltage-driven magnetic transition. Finally, Sec.~\ref{sec:conclusion} offers concluding remarks and an outlook.

\section{Machine learning architecture for conservative field}

\label{sec:BP}

\subsection{Spin dynamics of itinerant electron magnets}

We begin by formulating the governing equations for itinerant electron magnets. As a prototypical example, we consider the s-d exchange model, which captures the interaction between localized magnetic moments---typically originating from $d$ or $f$ electrons---and itinerant conduction electrons from the $s$ band. The general form of the single-band s-d exchange Hamiltonian can be written as
\begin{eqnarray}
	\label{eq:H1}
	\hat{\mathcal{H}} = -\sum_{ij, \alpha} t_{ij} \hat{c}^\dagger_{i\alpha} \hat{c}^{\,}_{j\beta} 
	- J \sum_{i, \alpha\beta} \mathbf {S}_i \cdot  \bm\sigma^{\,}_{\alpha\beta} \, \hat{c}^\dagger_{i\alpha} \hat{c}^{\,}_{i \beta},  \quad
\end{eqnarray}
where $\hat{c}^\dagger_{i\alpha}$ ($\hat{c}^{\,}_{i\alpha}$) denotes the creation (annihilation) operator for a conduction electron with spin $\alpha = \uparrow, \downarrow$ at lattice site $i$. The first term in Eq.~(\ref{eq:H1}) describes the Hamiltonian of the itinerant $s$ electrons, while the second term represents their local exchange interaction---also known as the Hund's-rule coupling---of strength $J$ with the localized magnetic moment $\mathbf{S}_i$ arising from the $d$ electrons. The electron operators in this term define the local spin density of the conduction electrons, $\hat{\mathbf{s}}_i = \frac{\hbar}{2}\bigl(\hat{c}^\dagger_{i\alpha}\bm{\sigma}_{\alpha\beta}\hat{c}^{\,}_{i\beta}\bigr)$. In most spintronic materials, the localized moments are sufficiently large that quantum fluctuations can be neglected, allowing them to be treated as classical spins of fixed magnitude $|\mathbf{S}_i| = S$. For convenience, the spin length is set to unity, $S = 1$, in the following.

The dynamics of the localized spins $\mathbf S_i$ are described by the Landau-Lifshitz-Gilbert (LLG) equation~\cite{LL,gilbert04}: 
\begin{eqnarray}
	\label{eq:LL}
	\frac{d \mathbf S_i }{ d t }= - \gamma \mathbf S_i \times \mathbf H_i
	+ \alpha  \mathbf S_i \times  \frac{d \mathbf S_i }{ d t } ,
\end{eqnarray} 
where $\gamma$ is the gyromagnetic ratio, $\alpha$ is an effective damping parameter, and $\mathbf H_i$ is a local magnetic field. In analogy with the molecular dynamics, this local electron-mediated exchange field can be viewed as a force acting on spin $\mathbf S_i$. For a conservative exchange field, which was considered in the original works of LL, this local field is given by 
\begin{eqnarray}
	\label{eq:H-force}
	\mathbf H_i = - \frac{\partial E}{\partial \mathbf S_i},
\end{eqnarray}
where $E = E(\{\mathbf S_i \})$ is the total energy of the system.  For small Hund's coupling $J_H \ll t$, by integrating out electrons using second-order perturbation theory, one obtains an effective interaction 
\begin{eqnarray}
	E = \frac{J^2}{W} \sum_{ij} \mathcal{J}(\mathbf r_i - \mathbf r_j) \, \mathbf S_i \cdot \mathbf S_j,
\end{eqnarray}
where $W$ is the bandwidth of the electron tight-binding model, and $\mathcal{J}(\mathbf r)$ is a (dimensionless) function which decays with the distance between the two spins, analogous to the well-known Ruderman-Kittel-Kasuya-Yosida (RKKY) interaction~\cite{Ruderman1954,Kasuya1956,Yosida1957}. For intermediate and strong Hund’s coupling, however, no simple perturbative form exists; instead, the effective energy required in the force calculation of Eq.~(\ref{eq:H-force}) must be obtained by integrating out the electrons on the fly:
\begin{eqnarray}
	\label{eq:E_system}
	E = \bigl\langle \hat{\mathcal{H}}(\{\mathbf S_i\}) \bigr\rangle 
	= {\rm Tr}\bigl(\hat{\rho}_e \hat{\mathcal{H}}  \bigr),
\end{eqnarray}
where $\hat{\rho}_e = \exp(- \hat{\mathscr{H}} / k_B T)/\mathcal{Z}$ is the equilibrium electron density matrix with respect to the instantaneous spin configurations, and $\mathcal{Z}$ is the partition function. For the s-d model in Eq.~(\ref{eq:H1}), the local field in the adiabatic approximation can be computed using the Hellmann-Feyman theorem~\cite{Hellmann37,Feynman39} $\mathbf H_i = -\partial \langle \hat{\mathcal{H}} \rangle / \partial \mathbf S_i = - \langle \partial \hat{\mathcal{H}} / \partial \mathbf S_i \rangle$, giving rise to
\begin{eqnarray}
	& & \mathbf H_i = J \sum_{\alpha\beta} \bm \sigma^{\,}_{\alpha\beta} \langle \hat{c}^\dagger_{j\beta} \hat{c}^{\,}_{i\alpha} \rangle \\
	& & \qquad = J \sum_m \sum_{\alpha\beta} \bm \sigma^{\,}_{\alpha\beta} f_{\rm FD}(\varepsilon_m) U^{(m)*}_{j\beta} U^{(m)}_{i\alpha}. \nonumber
\end{eqnarray}
The single-electron correlation function $\langle \hat{c}^\dagger_{j\beta} \hat{c}^{\,}_{i\alpha} \rangle$ can be directly computed from the eigenstates of the s-d Hamiltonian, as shown in the second equality above. Here $U^{(m)}_{i\alpha}$ is the eigenvector of the s-d Hamiltonian with eigen-energy $\varepsilon_m$. Alternatively, it can also be solved using the more efficient kernel polynomial method (KPM)~\cite{Weisse06,Barros13,Wang18}.


Quantum LLG simulations based on this framework have been used to investigate semiclassical spin-density-wave dynamics in Hubbard models~\cite{chern18} and anharmonic collective modes in double-exchange systems~\cite{bhattacharyya20}. As discussed in Sec.~\ref{sec:intro}, this quantum LLG approach closely parallels quantum MD~\cite{Marx09}: in both cases, the repeated evaluation of the electronic density matrix at each simulation step constitutes the dominant computational bottleneck. In the following, we introduce a scalable ML framework that eliminates this bottleneck and enables large-scale LLG dynamical simulations of s-d systems with near–quantum accuracy.

It is worth emphasizing that real materials generally also host short-range exchange interactions between local moments, which can be written as $E_{\rm ex} = \sum_{ij} \left( J_{ij}, \mathbf S_i \cdot \mathbf S_j + \mathbf D_{ij} \cdot \mathbf S_i \times \mathbf S_j \right)$. The corresponding contribution to the spin dynamics is straightforward to incorporate: the associated local field is obtained analytically as $\mathbf H^{\rm ex}_i = -\sum_j \left( J_{ij}\, \mathbf S_j + \mathbf D_{ij} \times \mathbf S_j \right)$, and can be added directly to the itinerant-electron field. Since the primary purpose of the ML framework developed here is to accelerate the computationally demanding evaluation of electron-mediated exchange fields, we restrict our attention throughout this paper to the s–d Hamiltonian.


\begin{figure*}[t]
\centering
\includegraphics[width=2\columnwidth]{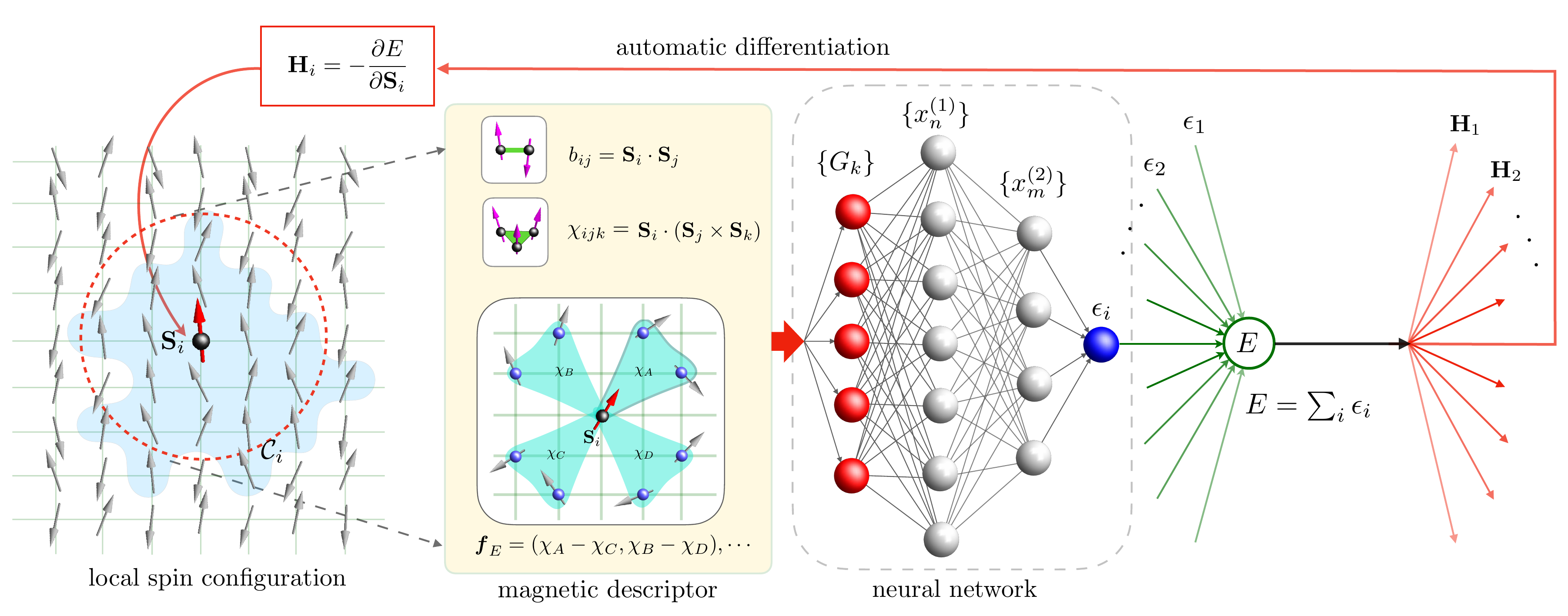}
\caption{Schematic diagram of ML force-field model for itinerant electron magnets. A descriptor transforms the neighborhood spin configuration $\mathcal{C}_i$ to effective coordinates $\{G_\ell \}$ which are then fed into a neural network (NN). The output node of the NN correspond to the local energy $\epsilon_i = \varepsilon(\mathcal{C}_i)$ associated with spin $\mathbf S_i$. $\{ x^{(1)}_m\}$ and $\{x^{(2)}_n\}$ denote feature variables associated with neurons of the first and second hidden layers, respectively.  The corresponding total potential energies $E$is obtained from summation of these local energies. Automatic differentiation is employed to compute the derivatives $\partial E / \partial \mathbf S_i$ from which the local exchange fields $\mathbf H_i$ are obtained. }
    \label{fig:BP-scheme}
\end{figure*}

\subsection{Behler-Parrinello ML architecture}

The ML force-field approach developed here provides an effectively classical force field, yet--owing to the remarkable expressive power of modern deep-learning models--it achieves accuracy comparable to full quantum calculations. To serve as a reliable surrogate for quantum dynamics, however, the ML framework must satisfy two key requirements: scalability and strict adherence to the underlying symmetry constraints. A central observation, originally emphasized by W. Kohn~\cite{kohn96}, is that true linear-scaling electronic-structure methods become possible only when the locality or “nearsightedness’’ principle holds~\cite{kohn96,prodan05}. This principle reflects the fact that, in many-electron systems, the influence of distant degrees of freedom is strongly suppressed by wave-mechanical destructive interference. Importantly, this form of locality is far more general than the existence of exponentially localized Wannier functions, which are restricted to large-gap insulators. Instead, nearsightedness is an intrinsic property of interacting quantum systems and governs the spatial decay of electronic responses even in metals and correlated materials.

Within this conceptual framework, ML force-field models provide an efficient and explicit way to harness locality. By encoding short-range electronic correlations directly into the ML architecture, these models implement the nearsightedness principle at the algorithmic level, enabling practical $\mathcal{O}(N)$ scaling in large-scale dynamical simulations. Indeed, the locality principle is the cornerstone of modern ML-based force-field approaches that now permit large-scale QMD simulations with the accuracy of {\em ab initio} density-functional theory (DFT) calculations~\cite{behler07,bartok10,li15,shapeev16,behler16,botu17,smith17,zhang18,deringer19,mcgibbon17,suwa19,chmiela17,chmiela18,sauceda20,unke21}. In these methods, atomic forces--central to QMD simulations--are assumed to depend primarily on the local chemical environment. A fixed-size ML model is then trained to capture the highly nontrivial mapping between an atom's local environment and the corresponding force acting upon it. Practical implementations of this paradigm were first demonstrated in the pioneering works of Behler and Parrinello~\cite{behler07} and Bart\'ok {\em et al.}~\cite{bartok10}. Similar ML frameworks have since been extended to a range of condensed-matter lattice systems, enabling large-scale modeling and simulation across diverse many-body settings~\cite{zhang20,zhang21,zhang22,zhang22b,zhang23,cheng23,cheng23b,Ghosh24,Fan24,Ma19,Liu22,Tian23}.

Here we generalize the Behler–Parrinello (BP) architecture~\cite{behler16} to describe the magnetization dynamics of metallic spin systems; a schematic illustration is shown in Fig.~\ref{fig:BP-scheme}. We begin by expressing the total energy $E$ in Eq.~(\ref{eq:E_system}) as a sum of local energy contributions,
\begin{eqnarray}
	\label{eq:decomp}
	E = \sum_i \epsilon_i = \sum_i \varepsilon\!\left(\mathcal{C}_i \right),
\end{eqnarray}
As indicated by the second equality, we invoke the locality principle and assume that the site energy $\epsilon_i$ depends only on a local magnetic environment, denoted as $\mathcal{C}_i$, through a universal function $\varepsilon(\cdot)$ which is determined by the underlying electron Hamiltonian. In practical implementations, the neighborhood is defined as the collection of spins within a cutoff radius~$r_c$ around the $i$-th site: $\mathcal{C}_i = \{ \mathbf S_j \,  \bigl| \, |\mathbf r_j - \mathbf r_i| < r_c \}$. The complex functional dependence of the local energy on the magnetic environment is to be approximated by a feed-forward neural network.

The feed-forward NN, shown in Fig.~\ref{fig:BP-scheme}, consists of a collection of artificial neurons, or nodes, arranged in successive layers. The input to the NN is the set of generalized coordinates describing the local spin environment, and the output is the corresponding local energy $\epsilon_i$. As information propagates from left to right, each neuron receives inputs ${x_i}$ from the previous layer and produces an output through a nonlinear transformation $y = F\left( \sum_i W_i x_i + b \right)$. Here $W_i$ and $b$ are the learnable weights and biases associated with each neuron, and $F(\cdot)$ is the activation function, which mimics the nonlinear, threshold-like response of a biological neuron. As a concrete illustration, consider the simplified architecture with two hidden layers shown in Fig.~\ref{fig:BP-scheme}. The local energy then takes the form
 \begin{eqnarray}
 	\label{eq:NN-func}
 	& & \epsilon_i =  \sum_m W^{(2)}_m x^{(2)}_m = \sum_m W^{(2)}_m F\biggl( \sum_k W^{(1)}_{mn} \, x^{(1)}_n + b^{(1)}_m \biggr) \quad\quad \\
	& & \quad = \sum_m W^{(2)}_m F\biggl( \sum_k W^{(1)}_{mn} F\biggl( \sum_k W^{(0)}_{n\ell} G_\ell + b^{(0)}_n \biggr) + b^{(1)}_m \biggr). \nonumber
 \end{eqnarray}
Here $x^{(1)}_n$ and $x^{(2)}_m$ denote the activations of neurons in the first and second hidden layers, and $\{G_\ell \}$ represents the set of input descriptors. The nested functional form on the right-hand side explicitly realizes the mapping $\varepsilon(\cdot)$ introduced in Eq.~(\ref{eq:decomp}), with the full set of weights ${W}$ and biases ${b}$ serving as trainable parameters that encode the local energy landscape. Thanks to the universal approximation theorem, such neural networks are, in principle, capable of representing unknown multidimensional real-valued functions with arbitrary accuracy~\cite{cybenko89,hornik89}.

The total energy $E$, given by the sum of local energies, is obtained by applying the same NN model to each individual spin in the lattice. This process is analogous to assembling identical ML models into a ``super neural network" that takes the entire spin configuration $\{ \mathbf S_i \}$ as input and outputs the total energy $E$ at its final node, as shown in Fig.~\ref{fig:BP-scheme}. The effective local field $\mathbf H_i$, defined as the derivative of the total energy as shown in Eq.~(\ref{eq:H1}), can be efficiently computed using automatic differentiation techniques~\cite{Paszke17,Baydin18}. In practice, however, due to the locality of the effective field, the calculation of $\mathbf H_i$ only requires contributions from local energies $\epsilon_j$ within a similar finite neighborhood around site-$i$

\section{Magnetic descriptor}

\label{sec:descriptor}

Another crucial component--already emphasized in the original works of Behler and Parrinello~\cite{behler07}--is the descriptors that faithfully represent the local chemical environment. Despite the powerful approximation capabilities of modern learning models, particularly deep neural networks, the symmetries of the underlying electronic Hamiltonian can only be captured statistically unless they are explicitly encoded in the input representation. For molecular systems, the relevant symmetry group is the three-dimensional Euclidean group $E(3)$, consisting of translations, rotations, and reflections, together with permutations of atomic species. Over the past decade, numerous atomic descriptors have been proposed~\cite{behler07,bartok10,li15,behler11,ghiringhelli15,bartok13,himanen20,huo22,drautz19}, including the widely used atom-centered symmetry functions (ACSF) and the group-theoretical bispectrum coefficients. More recently, it has been recognized that many of these descriptors arise as special cases of the Atomic Cluster Expansion (ACE) formalism~\cite{drautz19}, which provides a systematic, hierarchical, and in-principle complete framework for constructing symmetry-adapted representations of atomic environments.

In contrast, the situation for itinerant magnetic systems is far less developed. Atom-based descriptors designed for molecular structures cannot be directly applied to systems such as the s-d model, where additional symmetries associated with spin degrees of freedom must be respected. More broadly, descriptors for condensed-matter Hamiltonians containing internal variables--spins, orbitals, order parameters, or other field-like quantities--have yet to be systematically formulated. In such systems, the local environment must be represented in a manner consistent with both the spatial symmetries of the lattice and the symmetry operations acting on the internal degrees of freedom. Here we introduce a group-theoretical framework to construct a comprehensive descriptor for the s-d model---one that naturally incorporates these combined symmetries and can be generalized to a broad class of itinerant spin systems.

The s-d Hamiltonian in Eq.~(\ref{eq:H1}) is invariant under two independent symmetry groups: (i) global SO(3)/SU(2) spin rotations and (ii) the point group characterizing the lattice site symmetry. The rotation symmetry refers to simultaneous transformations of the classical local moments, $\mathbf S_i \to \mathcal{R}\cdot \mathbf S_i$, and the electron spinor operators, $\hat{c}_{i\alpha} \to \hat{U}_{\alpha\beta}\,\hat{c}_{i\beta}$, where $\mathcal{R}$ is an orthogonal $3\times 3$ rotation matrix and $\hat{U}=\hat{U}(\mathcal{R})$ is the corresponding $2\times 2$ unitary SU(2) operator acting on the electron spin.
The effective local-energy functional $\varepsilon(\mathcal{C}_i)$---obtained conceptually by integrating out the electrons?must therefore preserve the global SO(3) symmetry of the classical spins. This requirement can be satisfied manifestly by constructing the energy only from rotationally invariant spin combinations. The fundamental building blocks are the two-spin bond variable and the three-spin scalar chirality,
\begin{eqnarray}
	b_{jk} = \mathbf S_j \cdot \mathbf S_k, \qquad \chi_{jmn} = \mathbf S_j \cdot \mathbf S_m \times \mathbf S_n.
\end{eqnarray}
which encode two-body and three-body spin correlations, respectively. Any higher-order rotational invariants can be expressed as products and nonlinear combinations of these two quantities.

\begin{figure}
\includegraphics[width=0.99\columnwidth]{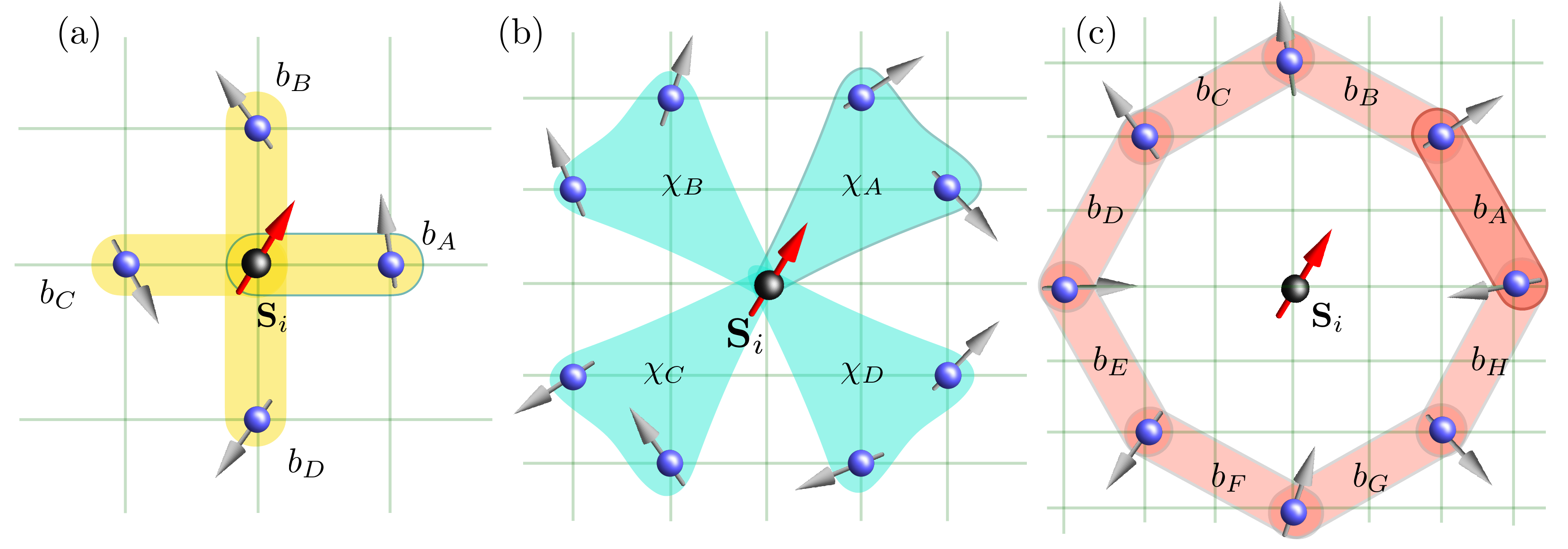}
\caption{(a) The four bond variables $b_A, b_B, b_C, b_D$, defined as the inner products between the central spin $\mathbf S_i$ and its four equidistant nearest neighbors, form the basis of a four-dimensional reducible representation of the $D_4$ point group. (b) The four scalar chiralities $\chi_A, \chi_B, \chi_C, \chi_D$ likewise span a four-dimensional reducible representation of $D_4$.  (c) The eight bond variables $b_A, b_B, \ldots, b_H$ constitute the basis of an eight-dimensional reducible representation.}
\label{fig:descriptor}  
\end{figure}

Next we examine the discrete lattice symmetries. Under any element of the lattice point group (rotations, reflections, inversion, etc.), a bond variable $b_{jk}$ is mapped to another bond $b_{j'k'}$ of the same bond length, and the chirality variables transform analogously. Consequently, the complete set of bond and chirality variables surrounding site $i$, ${ b_{jk},, \chi_{jmn} }$, furnishes the basis of a high-dimensional representation $\rho$ of the point group. To construct symmetry-invariant descriptors from these quantities, we first decompose this reducible representation into its irreducible components using standard group-theoretical procedures. Importantly, because point-group operations preserve the distance from the central spin at site $i$, the representation matrices of $\rho$ naturally acquire a block-diagonal structure when the variables are organized into shells of fixed radius. On the square lattice, these blocks have dimension 4 or 8, with some examples illustrated in Fig.~\ref{fig:descriptor}. The full representation matrix is therefore block-diagonal, which renders the subsequent decomposition particularly transparent.

As a concrete example, consider the four chirality variables in a size-4 block illustrated in Fig.~\ref{fig:descriptor}(b). Their irreducible components under the $D_{4}$ point group are
\begin{eqnarray*}
	& & f^{(A_1)} = \chi_A + \chi_B + \chi_C + \chi_D, \\
	& & f^{(B_1)} = \chi_A - \chi_B + \chi_C - \chi_D, \\
	& & f^{(E)}_1 = \chi_A - \chi_C, \quad f^{(E)}_2 = \chi_B - \chi_D,
\end{eqnarray*}
corresponding to the one-dimensional $A_1$ and $B_1$ representations and the two-dimensional $E$ representation, respectively.
Applying this decomposition procedure to all blocks, we obtain a set of basis vectors
\begin{eqnarray}
 \bm f^{(\Gamma, r)} = \bigl(f^{(\Gamma, r)}_1, f^{(\Gamma, r)}_2, \cdots, f^{(\Gamma, r)}_{n_\Gamma} \bigr)
\end{eqnarray}
where $\Gamma$ labels the irreducible representation, $r$ indexes its multiplicity, and $n_\Gamma$ is the dimension of the IR. These IR basis functions play a role analogous to Fourier components for the O(2) group or the spherical harmonics $Y_{\ell,m}$ for SO(3).
Importantly, the transformation of each vector $\boldsymbol{f}^{(\Gamma,r)}$ under a point-group operation is fully determined by the known matrix representation $\boldsymbol{D}^{(\Gamma)}$: $\bm f^{(\Gamma, r)} \to \bm D^{(\Gamma)} \cdot \bm f^{(\Gamma, r)}$. This property provides a systematic and rigorous route to constructing all point-group invariants, which in turn form the symmetry-adapted descriptors used in the neural-network representation of the local energy.

For example, analogous to ordinary Fourier analysis, the so-called power spectrum of irreducible representations (IRs) is invariant under the symmetry operations of the point group:
\begin{eqnarray}
	p^{(\Gamma, r)} = \bm f^{(\Gamma, r) \, \dagger} \cdot \bm f^{(\Gamma, r)} = \sum_{\mu = 1}^{n_\Gamma} \bigl| f^{(\Gamma, r)}_{\mu} \bigr|^2.
\end{eqnarray}
This invariance follows directly from the unitarity of the IR matrices, $\bm D^{(\Gamma),\dagger} \bm D^{(\Gamma)} = 1$, which represent the symmetry operations. However, descriptors based solely on the power spectrum possess artificial symmetries: they remain invariant even under independent transformations of individual IR vectors. What such descriptors fail to capture are the relative phases between different IR channels—quantities that are themselves invariant under the point group.
To illustrate, consider two doublet IRs, $\bm f^{(E,1)}$ and $\bm f^{(E,2)}$. Under point-group operations they each transform as 2D vectors. Hence, not only are their magnitudes $|\bm f^{(E,r)}|$ invariant, but so is their relative angle, encoded in $\cos\theta_{12} \propto \bm f^{(E,1)} \cdot \bm f^{(E,2)}$. Capturing this additional relational information requires extending beyond the power spectrum.

A complete symmetry-invariant descriptor is provided by the bispectrum coefficients~\cite{kondor07,bartok13},
\begin{eqnarray}
	\label{eq:bispectrum}
	b^{(\Gamma, \Gamma_1, \Gamma_2)}_{r, r_1, r_2} = \sum_{\kappa, \mu, \nu} C^{(\Gamma, \Gamma_1, \Gamma_2)}_{\kappa, \mu\nu} f^{(\Gamma, r) \, *}_\kappa f^{(\Gamma_1, r_1)}_\mu f^{(\Gamma_2, r_2)}_\nu, \quad
\end{eqnarray}
where $C^{(\Gamma,\Gamma_1,\Gamma_2)}_{\kappa,\mu\nu}$ are the Clebsch–Gordan coefficients of the point group. These bispectrum components are the group-theoretical analogs of scalar triple products among vectors and encode the relative phases between different IR channels.

For the D$_4$ point group—the symmetry group of the square lattice—the IRs are low-dimensional but exhibit large multiplicities (indexed by $r$), leading to an enormous number of potential bispectrum coefficients. Moreover, the full set of coefficients in Eq.~(\ref{eq:bispectrum}) is overcomplete. To keep the descriptor compact and computationally tractable, we introduce a reference-basis construction, $\bm g^{(\Gamma)}$, for each IR~$\Gamma$. These reference vectors are chosen to be insensitive to small variations in the local environment. For instance, one may construct a reducible representation from bond variables averaged over a large neighborhood and then project it onto the target IR to obtain $\bm g^{(\Gamma)}$.

Given these reference bases, the phase information for each IR channel can be encoded through simple projections, 
\begin{eqnarray}
	\eta^{(\Gamma,r)} = \bm g^{(\Gamma)} \cdot \bm f^{(\Gamma,r)}, 
\end{eqnarray}
while the relative phases between IRs are fully captured by the bispectrum coefficients constructed from the reference IRs, $b^{(\Gamma, \Gamma_1, \Gamma_2)}_{\rm ref} = \sum_{\kappa, \mu, \nu} C^{(\Gamma, \Gamma_1, \Gamma_2)}_{\kappa, \mu\nu} g^{(\Gamma) \, *}_\kappa g^{(\Gamma_1)}_\mu g^{(\Gamma_2)}_\nu$.

The feature variables of our descriptor are given by the combined set of amplitudes and relative phases associated with each IR,
\begin{eqnarray} \{G_\ell\} = \{ p^{(\Gamma, r)}, \eta^{(\Gamma,r)}, b^{(\Gamma,\Gamma_1, \Gamma_2)}_{\rm ref} \}. \end{eqnarray}
Crucially, these feature variables are invariant under both global SO(3) spin rotations and the D$4$ lattice symmetry operations. The full construction of the descriptor proceeds through the sequence
\begin{eqnarray}
	\{\mathbf S_j\} \,\, \to \,\, \{b_{jk}, \ \chi_{jmn} \} \,\, \to \,\, \{\bm f^{(\Gamma, r)}\} \,\, \to \,\, \{ G_{\ell}\}. \nonumber
\end{eqnarray}
where ${G_\ell}$ serves as a set of effective coordinates that compactly characterize the local spin environment. These symmetry-preserving coordinates form the input to the feed-forward neural network; see Fig.~\ref{fig:BP-scheme}. Since the local energy is expressed as $E_i = \varepsilon({G_\ell})$, depending exclusively on these invariant coordinates, the resulting ML energy function automatically respects all symmetries of the underlying Hamiltonian.


\section{ML models for non-collinear magnetic orders}

\label{sec:m-orders}

\begin{figure*}[t]
\includegraphics[width=1.8\columnwidth]{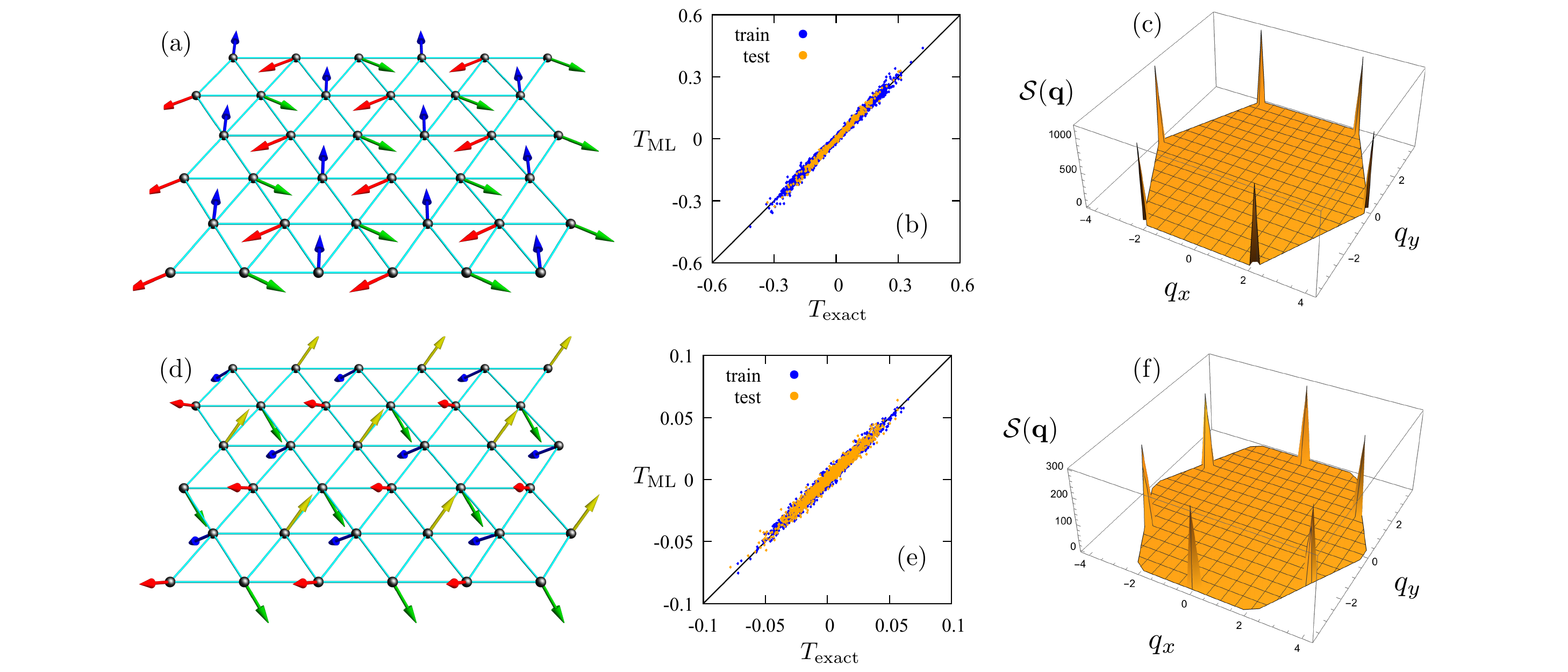}
\caption{Top panels: (a) The non-collinear 120$^\circ$ magnetic order with the tripled unit cell of the half-filled triangular-lattice s–d model. (b) Benchmark of ML-predicted torques, $\mathbf T_{\rm ML} = \mathbf S \times \mathbf H_{\rm ML}$, evaluated for the 120$^\circ$ state. (c) Ensemble-averaged spin structure factor $\mathcal{S}(\mathbf q)$ obtained from ML-LLG simulations. Bottom panels: Results for the non-coplanar tetrahedral order at filling fraction $f \approx 1/4$: (d) real-space spin configuration with a quadrupled magnetic unit cell; (e) ML torque-prediction benchmark; and (f) the corresponding ensemble-averaged spin structure factor.  }
\label{fig:ml-summary}  
\end{figure*}

As a first application, we employ the BP force-field model to the s-d system on a triangular lattice, where geometric frustration naturally promotes complex non-collinear magnetic orders. For the electronic Hamiltonian in Eq.~(\ref{eq:H1}), we use a nearest-neighbor tight-binding model. Given a classical spin configuration ${\mathbf S_i}$, the s-d Hamiltonian is solved via either exact diagonalization (ED) or the kernel polynomial method (KPM) to generate the training dataset. The triangular-lattice s-d model has been widely studied~\cite{Martin2008,Akagi10,Kato2010,Azhar17,Chern2012,Nandkishore2012}; the interplay of frustration and long-range electron-mediated interactions yields a rich phase diagram with non-collinear orders controlled by the Hund's coupling $J$ and electron filling.

To incorporate lattice symmetries, we construct descriptors that explicitly preserve the $D_6$ point-group symmetry of the triangular lattice~\cite{Fan24}. All neural-network models are implemented in PyTorch~\cite{Paszke2019} using fully connected architectures. The network consists of eight hidden layers with neuron counts $2048 \rightarrow 1024 \rightarrow 512 \rightarrow 256 \rightarrow 128 \rightarrow 64 \rightarrow 64 \rightarrow 64$. The input dimension is set to 1806, corresponding to the number of symmetry-preserving descriptor variables ${G_\ell}$ within a neighborhood of radius $r_c=6a$ and depth $l_c=2a$, where $a$ is the lattice constant. The output layer has a single neuron that predicts the local site energy $\epsilon_i$. Rectified Linear Unit (ReLU) activations~\cite{Nair2010} are used throughout.

Because the spin dynamics is drive by the torque $\mathbf T_i = \mathbf S_i \times \mathbf H_i$ in LLG equation, the loss function includes mean-square errors for both torques and total energy,
\begin{equation}
	\mathcal{L}(\bm \theta) = \sum^N_{i=1} \bigl| \mathbf T_i - \hat{\mathbf T}_i \bigr|^2
	+ \eta_E \Bigl| E - \sum_i^N \hat{\epsilon}_i \Bigr|^2. \,\,
	\label{eq:loss_func}
\end{equation}
where hatted quantities denote ML predictions dependent on trainable parameters $\boldsymbol\theta = \{ W, b\}$ of the neural network. Optimization is performed using the Adam algorithm~\cite{Kingma2014} with an exponentially decaying learning rate, starting from 0.1 and ending at 0.0001. The training set consists of 40,000 snapshots obtained from exact electronic-structure calculations. The model is trained for 200 epochs, and to reduce overfitting we employ 5-fold cross-validation~\cite{Stone1974}.

The trained ML force fields were incorporated into stochastic LLG simulations and benchmarked against reference simulations using KPM- or ED-computed local fields. We integrate the LLG equation using a standard semi-implicit scheme~\cite{Mentink10}. Because the spins have fixed length $|\mathbf S_i|=1$, the effective field carries units of energy, and the gyromagnetic ratio $\gamma$ has units of $1/\hbar$. In practice, we set $\gamma = 1$ and the nearest-neighbor hopping $t_{\rm nn}=1$, so time is measured in units of $\tau_0 = (\gamma t_{\rm nn})^{-1}$. Unless otherwise noted, all simulations use a timestep $\Delta t = 0.025\,\tau_0$ and a dissipation coefficient $\alpha = 0.075$.

We begin with the nearest-neighbor s-d model at half filling and strong Hund's coupling, $J = 6 t_{\rm nn}$, which serves as a benchmark for our ML force-field approach. In the limit $J \gg t_{\rm nn}$, the half-filled s-d model reduces to a classical antiferromagnetic Heisenberg model on any lattice~\cite{chern18}. At $t_{\rm nn}=0$, each site hosts one electron aligned with the local moment $\mathbf S_i$, yielding a massively degenerate manifold analogous to the large-$U$ half-filled Hubbard model~\cite{Fazekas99}. Treating the hopping perturbatively, the leading second-order process generates an effective exchange $E_{ij} = -2 (t^{\rm eff}_{ij})^2 / J$, where the effective hopping is $t^{\rm eff}_{ij} = t_{\rm nn} \sin(\theta_{ij} / 2)$. This leads to a Heisenberg Hamiltonian $E = J_{\rm eff}\sum_{\langle ij\rangle} \mathbf S_i\cdot \mathbf S_j$, with $J_{\rm eff}=4 t_{\rm nn}^2/J>0$. On the triangular lattice, this frustration stabilizes the 120$^\circ$ spin order with ordering wave vector $\mathbf K = (4\pi/3a,0)$.

To test our ML force field, we trained a neural network using 40,000 configurations obtained from KPM-based ED-LLG simulations on a $48\times48$ lattice. As shown in Fig.~\ref{fig:ml-summary}(a), the predicted torques agree extremely well with ground truth, reaching a MSE of $8.8\times10^{-8}$ with no noticeable overfitting. Using this model, we performed ML-LLG simulations of a thermal quench on a $150\times150$ system. The late-time state [Fig.~\ref{fig:ml-summary}(b)] exhibits the expected three-sublattice 120$^\circ$ pattern. This is further confirmed by the computation of structure factor defined as
\begin{eqnarray}
	\mathcal{S}(\mathbf q, t) = \left\langle \Bigl| \frac{1}{N} \sum_i \mathbf S_i(t) e^{i \mathbf q \cdot \mathbf r_i} \Bigr|^2  \right\rangle
\end{eqnarray}
where $\langle \cdots \rangle$ denotes ensemle average. As shown in Fig.~\ref{fig:ml-summary}(c), the structure factor obtained from ML-LLG shows clear Bragg peaks at the six BZ corners corresponding to the ordering wavevector $\mathbf K$.

We next turn to the noncoplanar tetrahedral order, the ground state of the triangular-lattice s-d model near fillings $f=3/4$ and $f\sim1/4$~\cite{Martin2008,Akagi10,Kato2010,Azhar17,Chern2012}. This state is a triple-$\mathbf Q$ structure, 
\begin{eqnarray}
	\mathbf S_i = \sum_{\eta = 1, 2, 3} \mathbf {\bm \Delta}_{\eta} e^{i \mathbf M_\eta \cdot \mathbf r_i},  
\end{eqnarray}
with $\mathbf M_1=(2\pi/a,0)$ and $\mathbf M_{2,3}=(-\pi/a,\pm\sqrt{3}\pi/a)$. Choosing orthogonal $\bm\Delta_\eta$ of equal amplitude produces the highly symmetric tetrahedral configuration on a quadrupled unit cell.
At $3/4$ filling, the tetrahedral order arises from perfect Fermi-surface nesting~\cite{Martin2008,Chern2012}. Near $1/4$ filling, however, the mechanism is different: although the Lindhard susceptibility $\chi(\mathbf q)$ computed to second order in $J$ shows only weak maxima at $\mathbf M_\eta$, the triple-$\mathbf Q$ degeneracy is lifted at fourth order. A positive biquadratic term $B(\mathbf S_i\cdot\mathbf S_j)^2$ ($B \sim J^4/t_{\rm nn}^3>0$) favors the tetrahedral state~\cite{Akagi2012}. This enhancement originates from a Fermi-surface geometry in which the three wave vectors $\mathbf M_\eta$ connect nearly parallel-tangent points, producing a singularity analogous to the $2k_F$ Kohn anomaly.

To model this regime, we trained an ML force field for $J/t_{\rm nn}=3$ using KPM data generated in the grand-canonical ensemble with chemical potential $\mu=-3.2 t_{\rm nn}$ (near quarter filling). The force prediction benchmark [Fig.~\ref{fig:ml-summary}(d)] yields a MSE of $7.8\times10^{-7}$ without overfitting; additional dynamical benchmarks are given in Ref.~\cite{Fan24}. ML-LLG quenches produce a clear four-sublattice structure with small fluctuations [Fig.~\ref{fig:ml-summary}(e)], and the corresponding structure factor [Fig.~\ref{fig:ml-summary}(f)] shows sharp peaks at the six $M$ points, confirming the triple-$\mathbf Q$ tetrahedral ordering.

It is worth noting that the tetrahedral state also breaks a $Z_2$ chiral symmetry, reflected in a nonzero scalar chirality $\chi_{ijk}=\mathbf S_i\cdot(\mathbf S_j\times\mathbf S_k)$ on each triangular plaquette. This noncoplanar order yields a quantized Hall response of $\pm e^2/h$, with the sign set by the chirality~\cite{Martin2008,Chern2012}. Although long-range tetrahedral order is thermally unstable in two dimensions, the $Z_2$ chiral order---an Ising-like degree of freedom---remains robust at finite temperatures. Remarkably, our ML-enabled large-scale LLG simulations reveal a linear growth of domain size~\cite{Fan24}, in sharp contrast to the Allen-Cahn $t^{1/2}$ law expected for a nonconserved Ising order parameter~\cite{Bray1994,Puri2009,Onuki2002}. This accelerated growth originates from the strong orientational anisotropy of the chiral domain boundaries.

\section{Relaxation dynamics of mixed-phase states}

\label{sec:relaxation}

As a second application, we use the ML force-field framework to investigate the relaxation of complex spin textures that emerge in the mixed-phase states of the s-d model on a square lattice~\cite{zhang20,zhang21}. In the strong-coupling limit—often referred to as the double-exchange regime—the s-d model develops a thermodynamically stable mixed-phase state when doped slightly away from half-filling~\cite{zener51,anderson55,degennes60}. Phase separation of this type is a common feature of strongly correlated electron systems and often plays a central role in determining their macroscopic transport and magnetic responses~\cite{schulz89,emery90,white00,tranquada95,dagotto_book,dagotto05,moreo99,dagotto01,mathur03,nagaev02}.

\begin{figure}
\includegraphics[width=0.99\columnwidth]{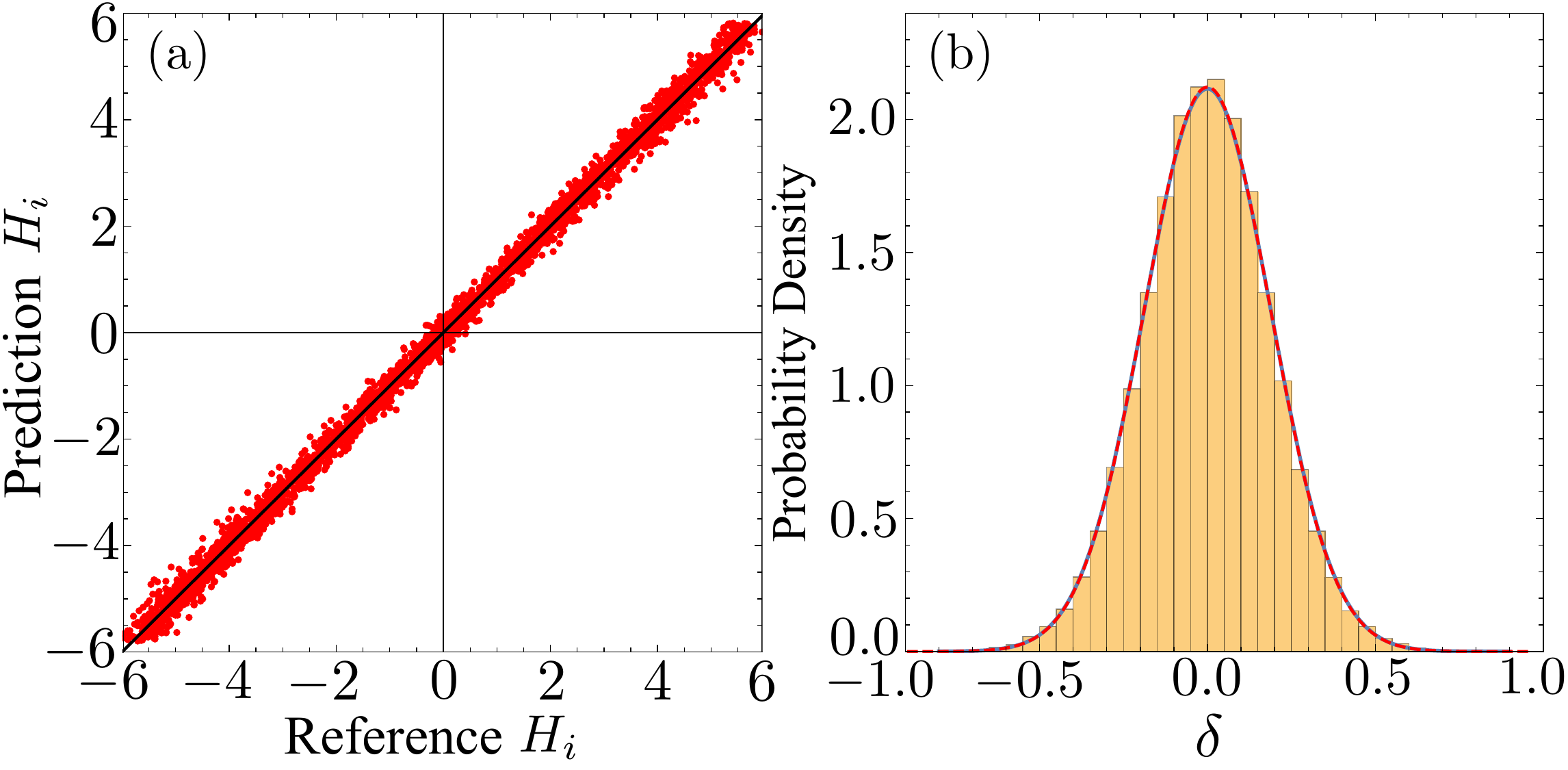}
\caption{(a) ML-predicted exchange fields compared against exact values from the test dataset. (b) Distribution of the force difference $\delta = H_{\text{ML}} - H_{\text{exact}}$, which is well described by a normal distribution (red curve) with variance $\sigma^2 = 0.035$. 
\label{fig:ed-force} 
}
\end{figure}

A prototypical example is found in manganites and related magnetic semiconductors that exhibit colossal magnetoresistance (CMR). Across a wide doping range, these materials host intricate nanoscale textures~\cite{dagotto_book,dagotto05,moreo99,dagotto01,mathur03,nagaev02} consisting of hole-rich ferromagnetic (FM) clusters embedded in nearly half-filled antiferromagnetic (AFM) regions. Scanning-probe and spectroscopic techniques have provided direct evidence for such spatially inhomogeneous states~\cite{fath99,renner02,salamon01}. The s-d model captures the essential physics of this competition and therefore serves as an ideal testing ground for ML-driven simulations of mixed-phase dynamics.

Because the carrier density is conserved, the evolution of these inhomogeneous states is expected to follow the classical Lifshitz-Slyozov-Wagner (LSW) theory of conserved-order-parameter coarsening~\cite{lifshitz61,wagner61}. In such systems, domain growth is controlled by the diffusive redistribution of doped holes, leading to the characteristic scaling law $L(t) \sim t^\alpha$ with the universal exponent $\alpha = 1/3$. The double-exchange s-d model thus provides a natural platform for examining whether ML-based force-field models faithfully capture mesoscale phase separation, dynamical scaling, and late-time coarsening kinetics.

The static and thermodynamic properties of the square-lattice double-exchange model have been extensively investigated~\cite{varma96,yunoki98,dagotto98,chattopadhyay01,pekker05}. At small hole doping, a mixed-phase state comprising FM puddles embedded in an AFM insulating background emerges as a robust ground state at strong Hund’s coupling~\cite{yunoki98,dagotto98,chattopadhyay01}. Despite this detailed understanding of the equilibrium textures, the nonequilibrium kinetics of correlation-driven phase separation—particularly the scaling behavior and the mechanism of late-stage growth—remain largely unexplored. The scalability and efficiency of the ML force-field approach enable large-scale dynamical simulations capable of addressing these open questions.

To construct the ML force field, we trained a fully connected six-layer neural network implemented in PyTorch~\cite{Paszke2019,Nair2010} using 3500 snapshots of spin configurations and local exchange fields obtained from ED–LLG simulations on a $30\times30$ lattice. Fig.~\ref{fig:ed-force}(a) compares the NN-predicted exchange fields $\mathbf H_i$ with exact values from ED, while Fig.~\ref{fig:ed-force}(b) shows that the deviation $\delta = H_{\text{ML}} - H_{\text{exact}}$ is well-described by a Gaussian distribution with variance $\sigma^2 = 0.035$. The approximately normal distribution of $\delta$ suggests that ML uncertainty effectively acts as a small stochastic field, akin to a thermal noise term in Langevin dynamics. 

\begin{figure*}
\includegraphics[width=1.98\columnwidth]{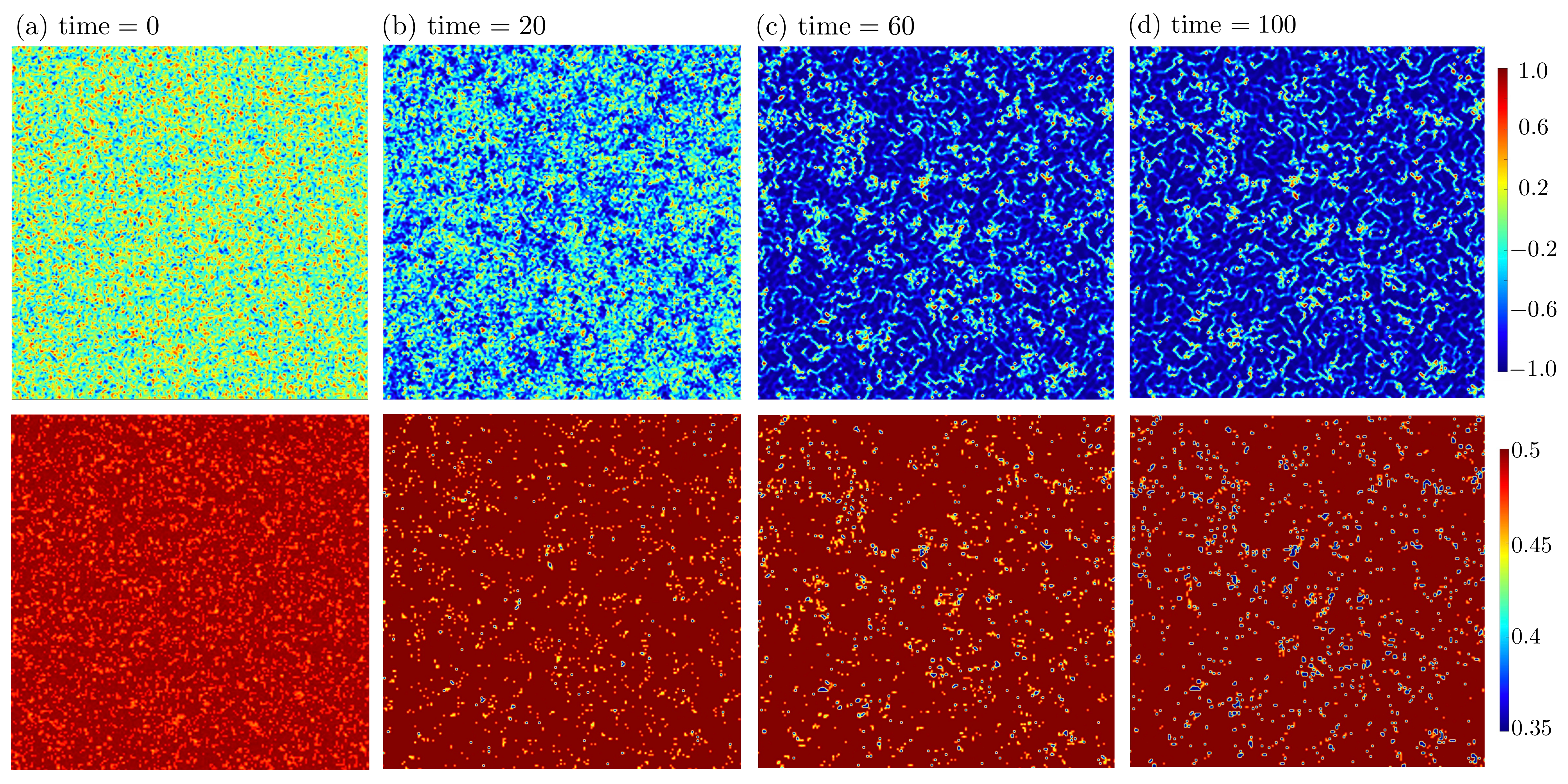}
\caption{(Top) Density plots of the local nearest-neighbor spin correlation $b_i \equiv (\mathbf S_i \cdot \mathbf S_{i+\mathbf x} + \mathbf S_i \cdot \mathbf S_{i-\mathbf x} + \mathbf S_i \cdot \mathbf S_{i+\mathbf y} + \mathbf S_i \cdot \mathbf S_{i-\mathbf y})/4$ at four representative times during the ML-LLG simulation on a $200 \times 200$ lattice. (Bottom) Corresponding ML-predicted distributions of the local electron number $n_i = \sum_\alpha \langle \hat{c}^\dagger_{i\alpha} \hat{c}^{\,}_{i\alpha} \rangle$. The neural-network model is trained on datasets obtained from exact solutions for an $L = 30$ lattice with parameters $t{\rm nn} = 1$, $J = 7$, and electron filling fraction $n = 0.485$. 
\label{fig:snapshots} 
}
\end{figure*}

We then combined the trained network with LLG time evolution to simulate quenches from random initial configurations at $T = 0.022$ and filling $f = 0.485$ (1.5\% hole doping). Using the trained model, we carried out large-scale quench simulations on a $200\times200$ lattice. The top panels of Fig.~\ref{fig:snapshots} shows snapshots of the local bond correlation $b_i = (\mathbf S_i \cdot \mathbf S_{i+\mathbf x} + \mathbf S_i \cdot \mathbf S_{i-\mathbf x} + \mathbf S_i \cdot \mathbf S_{i+\mathbf y} + \mathbf S_i \cdot \mathbf S_{i-\mathbf y})/4$ during the relaxation process of a thermal quench. Here positive (negative) values of $b_i$ indicate FM (AFM) regions. The ML-LLG dynamics relaxes into an inhomogeneous configuration consisting of extended AFM domains interspersed with small FM droplets characteristic of weakly doped double-exchange systems.

The double-exchange mechanism implies a strong correlation between spin order and carrier density: FM regions tend to coincide with doped holes, whereas the AFM background is favored at or near exact half-filling. To examine this relation more quantitatively, we trained a separate NN model to directly predict the on-site electron density $n_i = \sum_\alpha \langle \hat{c}^\dagger_{i\alpha} \hat{c}^{\,}_{i\alpha} \rangle$ from the neighborhood spin configuration $\mathcal{C}_i$. As before, a symmetry-preserving descriptor is used to map $\mathcal{C}_i$ onto a set of invariant features $\{G_{\ell} \}$, which serve as inputs to the ML density model.
As shown in the bottom panels of Fig.~\ref{fig:snapshots}, doped holes cluster into small puddles characterized by predominantly FM spin correlations embedded within the AFM background at half-filling. Interestingly, the coarsening of these hole puddles appears to be quite sluggish, with their characteristic sizes remaining relatively small even at late times.

\begin{figure}
\includegraphics[width=0.99\columnwidth]{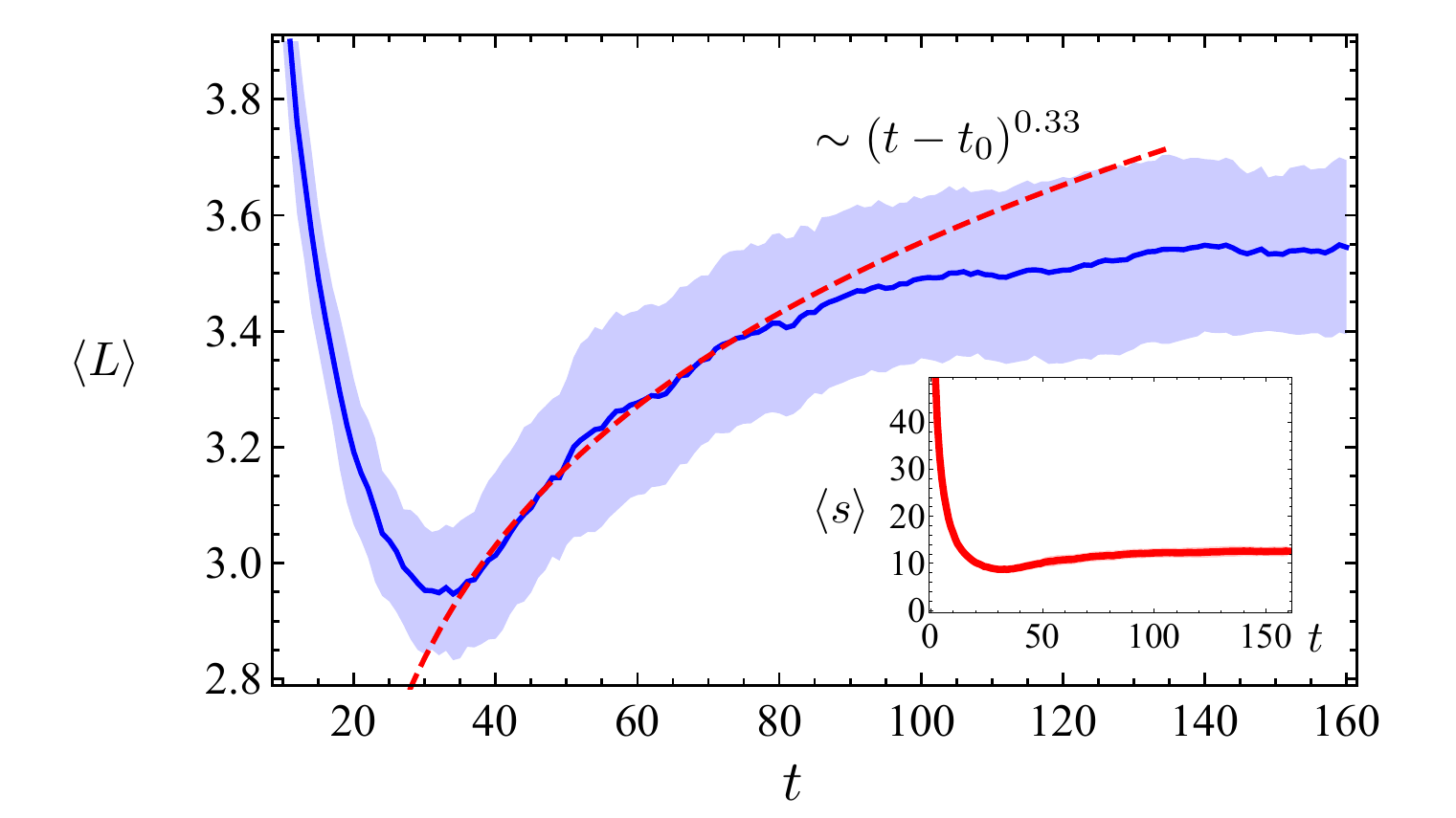}
\caption{Average linear size $L = \langle s \rangle^{1/2}$ of ferromagnetic (FM) clusters as a function of time following a thermal quench. The dash–dot line marks the power-law growth $L(t) \sim t^{1/3}$, while the dashed line shows a sublogarithmic form $L(t) \sim (\log t)^{\beta}$ with $\beta = 0.11$. The inset displays the corresponding time evolution of the average cluster size $\langle s \rangle$.
\label{fig:relaxation}
}
\end{figure}

To quantify the growth kinetics of hole puddles associated with FM spin correlations, we define FM clusters as connected regions where all nearest-neighbor bonds satisfy $b_{\rm th} > 0.5$. The evolution of the characteristic domain size $L(t)$ is shown in Fig.~\ref{fig:relaxation}. As expected for a conserved-order-parameter system, the early-time growth follows the LSW prediction $L(t) \sim t^{1/3}$. However, at later times the coarsening slows markedly, and $L(t)$ deviates from the LSW power law. Classical LSW theory describes growth through an evaporation-condensation mechanism in which larger domains grow by absorbing material from smaller ones via diffusive exchange. In the present context, this process corresponds to the migration of doped holes from small FM clusters into larger ones. The initial $t^{1/3}$ scaling likely reflects this early diffusive regime. As AFM order strengthens in the half-filled background, each doped hole becomes surrounded by parallel spins via the double-exchange mechanism, forming a self-trapped FM cloud. This localization suppresses hole evaporation, arrests further coarsening, and ultimately leads to a breakdown of the LSW scenario at late times.

\section{Non-conservative field and Generalized BP architecture}

\label{sec:generalized-BP}

The BP-type ML architecture developed in the preceding sections enables linear-scaling, adiabatic dynamical simulations of large itinerant spin systems. Within the adiabatic approximation, the electronic subsystem is assumed to remain in quasi-equilibrium with the instantaneous spin configuration throughout the evolution. Consequently, the system energy in Eq.~(\ref{eq:E_system}) is evaluated using the Boltzmann distribution of the electron liquid corresponding to the current spin state. In this framework, the local magnetic fields $\mathbf H_i$ that govern the spin dynamics are obtained by differentiating the total energy constructed from the ML-predicted local energies. This procedure is made computationally efficient through the use of automatic differentiation techniques.

However, the fact that forces are obtained as derivatives of the predicted energy also underscores a fundamental limitation of the BP framework: it is intrinsically restricted to representing conservative forces. Exchange fields generated by out-of-equilibrium electrons or in open systems are generally non-conservative and therefore cannot, in general, be written as the gradient of any effective potential. In such settings, even the notion of a well-defined total energy $E$ becomes ambiguous. Capturing these forces requires ML representations that go beyond the BP architecture.

A well-known example is the current-induced force in molecular junctions~\cite{lu12,todorov10,dundas09,diventra00}, which has been explicitly shown to be nonconservative. Another important case is the spin-transfer torque~\cite{slonczewski96,berger96,brataas12,ralph08,salahuddin08} produced by polarized charge currents, a key mechanism in nanomagnetism and spintronics. These examples highlight the broader challenge: it remains unclear how the extensive ML machinery developed for quasi-equilibrium QMD can be directly adapted to model electronic dynamics far from equilibrium, where forces are neither conservative nor derived from an underlying potential landscape.

We address this challenge in quantum LLG dynamics by showing that general nonconservative forces in itinerant magnets can be represented using two scalar potentials~\cite{zhang23}. This reformulation reduces force prediction to learning two local energies. Using the locality principle, we construct a generalized BP architecture that predicts these energies, with forces obtained via automatic differentiation. The scalar outputs naturally incorporate symmetry constraints and retain the advantages of BP-type models. As a demonstration, we train the framework on nonequilibrium exchange fields from Green’s-function calculations for the s–d model and show that it accurately reproduces voltage-driven domain-wall motion.

\begin{figure}
\includegraphics[width=1.0\columnwidth]{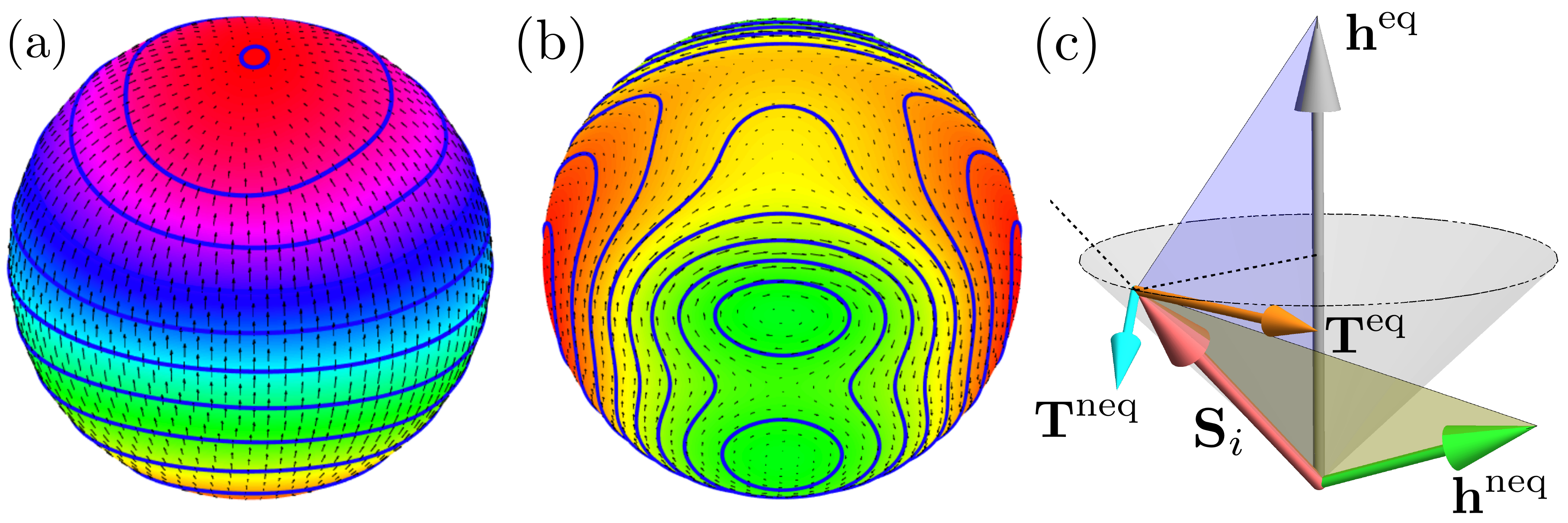}
\caption{A tangential vector field on a sphere can be decomposed into (a) the curl-free component $\nabla_s \,{E}$ and (b) the divergence-free component $\nabla_s \times {G}$. (c) shows the gradient field $\mathbf h^{\rm eq}_i = -\partial {E}/\partial \mathbf S_i$, which can be viewed as a quasi-equilibrium exchange field, and the curl-field $\mathbf h^{\rm neq}_i = -\mathbf S_i \times \partial {G} /\partial \mathbf S_i$, which corresponds to the nonequilibrium force, and their respective torques $\mathbf T^{\rm eq}_i = \mathbf h^{\rm eq}_i \times \mathbf S_i$ and $\mathbf T^{\rm neq}_i = \mathbf h^{\rm neq}_i \times \mathbf S_i$.}
\label{fig:decomposition} 
\end{figure}

\begin{figure*}
\includegraphics[width=1.99\columnwidth]{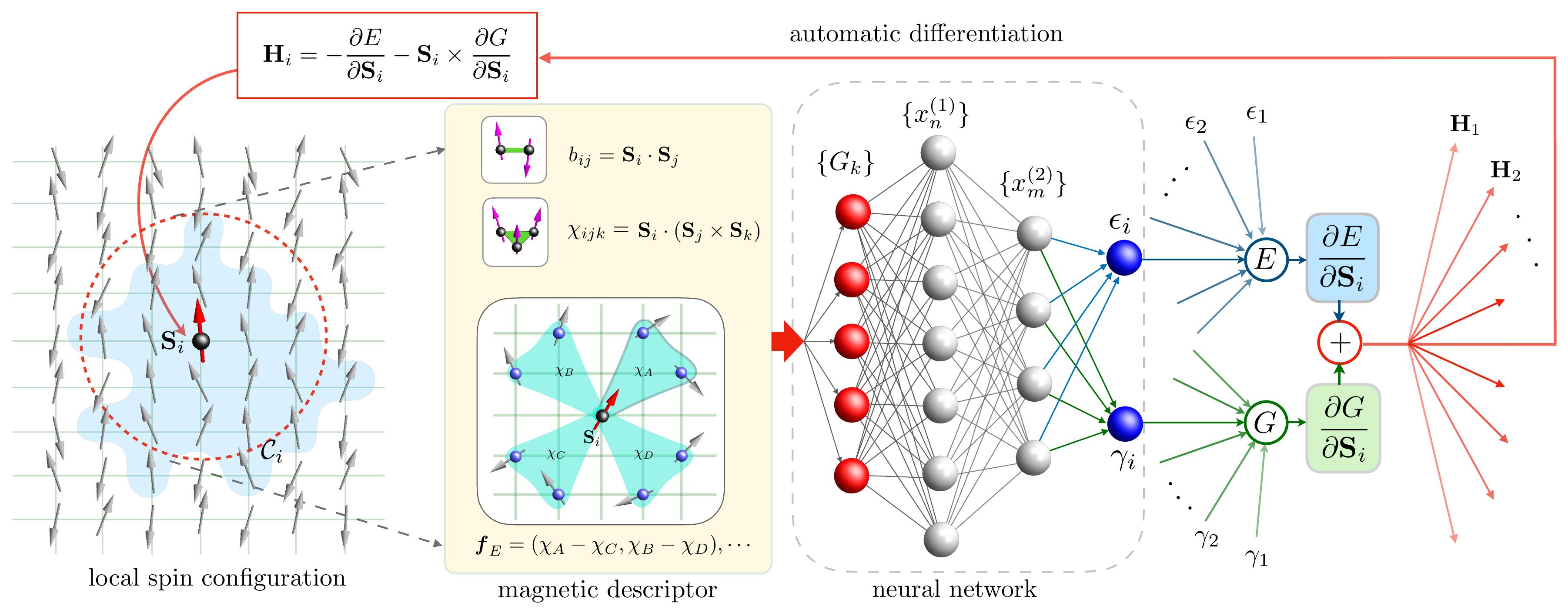}
\caption{(a) Schematic diagram of neural-network (NN) force-field model for out-of-equilibrium itinerant spin system. A descriptor transforms the neighborhood spin configuration $\mathcal{C}_i$ to effective coordinates $\{G_\ell\}$ which are then fed into a NN.  The two output nodes of the NN correspond to the local energy~$\epsilon_i = \varepsilon(\mathcal{C}_i)$ and $\gamma_i = \chi(\mathcal{C}_i)$ associated with site-$i$. The corresponding total potential energies ${E}$ and ${G}$ are obtained from summation of these local energies. Automatic differentiation is employed to compute the derivatives $\partial {E} / \partial \mathbf S_i$ and $\partial {G} / \partial \mathbf S_i$, from which the local exchange fields~$\mathbf H_i$ are obtained according to the generalized force expression~Eq.(\ref{eq:force}).
}
\label{fig:generalized-BP} 
\end{figure*}

\subsection{Generalized potentials for LLG}

A defining feature of the LLG equation is the strict conservation of spin length. Consequently, the effective field $\mathbf H$ in Eq.~(\ref{eq:LL}) should be regarded as a vector field on the sphere $S^2$ defined by $|\mathbf S_i| = S$. By the Helmholtz-Hodge decomposition on $S^2$~\cite{adams62,swarztrauber81,fan18}, any such field $\mathbf H(\mathbf S)$ can be uniquely decomposed into radial, gradient, and solenoidal components. The radial part, being parallel to $\mathbf S$, produces no torque $\mathbf T = \mathbf S \times \mathbf H$ and therefore behaves as a gauge freedom with no physical effect on the dynamics. The physically relevant exchange field thus consists only of the tangential gradient and solenoidal contributions,
\begin{eqnarray}
	\mathbf H = - \nabla_s\, {E}(\mathbf S) - \nabla_s \times {G}(\mathbf S),
\end{eqnarray}
where ${E}$ and ${G}$ are scalar functions on the spin sphere. The surface gradient of a scalar $f(\mathbf S)$ is
\begin{eqnarray}
	\nabla_s f = \frac{\partial f}{\partial \mathbf S} - \mathbf S \left( \mathbf S \cdot \frac{\partial f}{\partial \mathbf S} \right),
\end{eqnarray}
and the corresponding surface curl is
\begin{eqnarray}
	\nabla_s \times f = \mathbf S \times \frac{\partial f}{\partial \mathbf S}.
\end{eqnarray}
Here $\frac{\partial f}{\partial \mathbf S} = \sum_{\alpha = x,y,z} \frac{\partial f}{\partial S^\alpha}\,\hat{\mathbf e}_\alpha$ denotes the ordinary three-dimensional gradient, prior to imposing the constraint $|\mathbf S|=\text{const}$. Notably, unlike the familiar Helmholtz decomposition in Euclidean space, $\mathbf V = -\nabla \phi + \nabla \times \mathbf A$, the curl component on $S^2$ is generated by a scalar potential~$G$. 

Since the second term in the surface gradient is parallel to the spin direction, it does not contribute to the torque that drives LLG dynamics. Thus, the most general exchange field in the LLG equation can be written in terms of two scalar potentials as
\begin{eqnarray}
\label{eq:force}
	\mathbf H_i = - \frac{\partial {E}}{\partial \mathbf S_i} - \mathbf S_i \times \frac{\partial {G}}{\partial \mathbf S_i}
	= \mathbf h_i^{\rm eq} + \mathbf h_i^{\rm neq}.
\end{eqnarray}
The first term, analogous to a conservative force, defines the quasi-equilibrium exchange field, while the second term, arising from the curl component, is the nonequilibrium exchange field; see Fig.~\ref{fig:decomposition}. The generalized LLG equation then becomes~\cite{zhang23}
\begin{eqnarray}
\label{eq:LL2}
	\frac{\partial \mathbf S_i }{ \partial t } = \gamma\, \mathbf S_i \times \frac{\partial {E}}{\partial \mathbf S_i}
	+ \gamma\, \mathbf S_i \times \left( \mathbf S_i \times \frac{\partial {G}}{\partial \mathbf S_i} \right)
	+ \alpha\, \mathbf S_i \times \frac{\partial \mathbf S_i }{ \partial t }. \nonumber \\
\end{eqnarray}
The first term describes the conventional precessional dynamics of Eq.~(\ref{eq:LL}), with ${E}$ serving as an effective conservative potential. The third term, proportional to the Gilbert damping constant~$\alpha$, accounts for the universal dissipation that monotonically decreases ${E}$. In contrast, the intermediate (toroidal) $G$ term can represent either energy loss or energy injection, depending on the choice of $G$. For example, setting ${G} = -\lambda {E}$ with $\lambda > 0$ reproduces the dissipative contribution introduced in the original formulation of Landau and Lifshitz~\cite{LL}. Conversely, nonequilibrium spin-transfer torques of the Slonczewski–Berger type~\cite{slonczewski96,berger96} arise from a potential of the form $G = - K \sum_i \mathbf M_p \cdot \mathbf S_i$, where $K$ is an effective exchange coupling and $\mathbf M_p$ denotes the magnetization of the fixed polarizing layer in a magnetic tunnel junction. Although this interaction resembles a Zeeman coupling to an external magnetic field, the resulting torque, $\mathbf h_i^{\rm neq} = K \mathbf S_i \times (\mathbf S_i \times \mathbf M_p)$ drives the local spin toward alignment with $\mathbf M_p$, rather than producing the precessional motion characteristic of a magnetic-field-induced Zeeman term.

\subsection{Generalized Behler-Parrinello architecture}

The above formulation of the generalized LLG dynamics in terms of two {\em scalar} potentials provides a natural pathway for extending the Behler-Parrinello (BP) architecture to represent non-conservative exchange fields. As in the standard BP scheme, this generalized approach retains the important advantage of incorporating the symmetry constraints of the underlying electronic Hamiltonian directly into the ML model through an appropriately designed magnetic descriptor. A schematic illustration of this generalized BP framework is shown in Fig.~\ref{fig:generalized-BP}.
To construct such a model, we begin by decomposing the two scalar potentials into sums of local contributions, namely ${E} = \sum_i \epsilon_i$ and ${G} = \sum_i \gamma_i$.
Invoking the principle of locality~\cite{kohn96,prodan05}, the corresponding local energies $\epsilon_i$ and $\gamma_i$ are assumed to depend solely on the local magnetic environment $\mathcal{C}_i$ through two universal (model-dependent) functions. That is, for a given electronic Hamiltonian, we posit $\epsilon_i = \varepsilon(\mathcal{C}_i)$ and $\gamma_i = \chi(\mathcal{C}_i)$. The resulting dependence of the total potentials on the full spin configuration ${\mathbf S_i}$ can thus be written as
\begin{eqnarray}
	\label{eq:EG}
	& & {E} = \sum_i \epsilon_i = \sum_i \varepsilon(\mathcal{C}_i), \nonumber \\
	& & {G} = \sum_i \gamma_i = \sum_i \chi(\mathcal{C}_i).
\end{eqnarray}
Similar to the quasi-equilibrium BP model discussed in Sec.~\ref{sec:BP}, the magnetic environment $\mathcal{C}_i$ is defined as the set of spins located within a cutoff radius $R_c$ of the $i$-th site,
$\mathcal{C}_i = \bigl\{ \mathbf S_j \, \big| \, |\mathbf r_j - \mathbf r_i| \le R_c \bigr\}$. The complicated functional dependences of the local energies on $\mathcal{C}_i$ are then approximated by a deep-learning neural network, as depicted in Fig.~\ref{fig:generalized-BP}. Consistent with the construction of the original BP architecture, the combined SO(3) spin-rotation symmetry and the lattice point-group symmetries are encoded through the magnetic descriptor introduced in Sec.~\ref{sec:descriptor}. The resulting symmetry-invariant features $\{ G_\ell \}$ are fed into a fully connected neural network that outputs the pair of local energies $\epsilon_i$ and $\gamma_i$ for each site.

Once all local contributions are generated by the NN, the full scalar potentials ${E}$ and ${G}$ follow directly from Eq.~(\ref{eq:EG}). The corresponding local exchange fields $\mathbf H_i$ entering the generalized LLG equation are then computed from the functional derivatives of these potentials according to Eq.~(\ref{eq:force}). These derivatives, $\partial {E} / \partial \mathbf S_i$ and $\partial {G} / \partial \mathbf S_i$, are evaluated both efficiently and accurately using modern automatic differentiation methods~\cite{Paszke17,Baydin18}. This procedure yields a unified, symmetry-preserving ML architecture capable of modeling both conservative and non-conservative components of the exchange field within a single generalized BP framework.

\section{Voltage-induced magnetic phase transition}

\label{sec:noneq}

\begin{figure}[t]
\includegraphics[width=0.95\columnwidth]{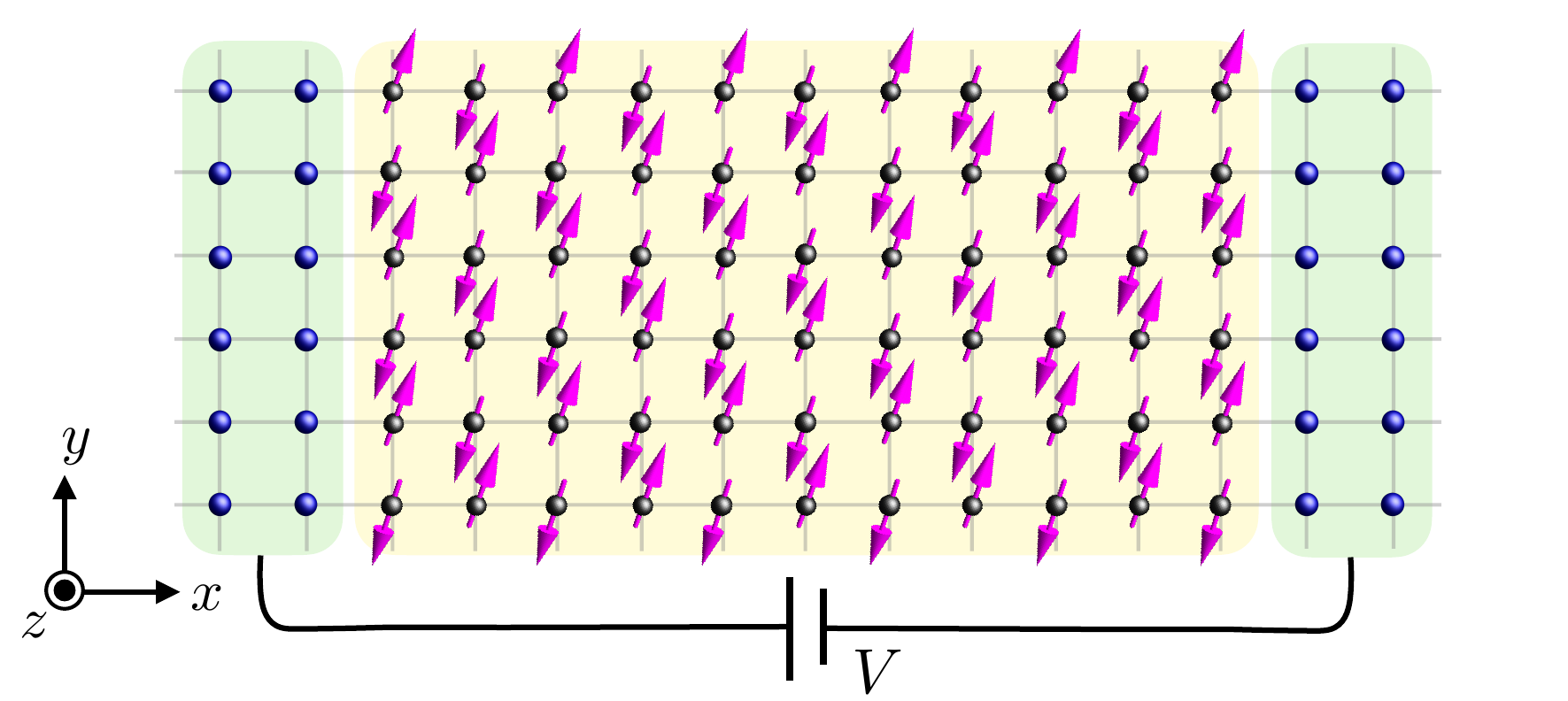}
\caption{(Color online)  
\label{fig:negf} Schematic diagram of the voltage-driven magnetic transition in the DE model sandwiched by two electrodes. In the initial state, spins in the DE model are arranged in a staggered N\'eel order. A voltage drop $V$ is applied to the two electrodes. 
}
\end{figure}

The ML framework developed above is fully general and can be applied to represent exchange fields in a wide range of nonequilibrium electronic systems. As a concrete demonstration, we apply it here to learn the forces obtained from nonequilibrium Green’s function (NEGF) calculations~\cite{datta95,haug08,diventra08} for a voltage-driven s-d model. We consider a square-lattice s-d system coupled to two metallic electrodes arranged in a capacitor-like geometry, as illustrated in Fig.~\ref{fig:negf}. The total Hamiltonian is written as $\mathcal{H}{\rm tot} = \mathcal{H}{\scriptsize \mbox{s-d}} + \mathcal{H}{\rm res}$, where $\mathcal{H}{\scriptsize \mbox{s-d}}$ is given in Eq.~(\ref{eq:H1}), and $\mathcal{H}_{\rm res}$ describes the electronic degrees of freedom in the electrodes and reservoirs, as well as their coupling to the central s-d region. Within the NEGF formalism, the effects of the reservoir fermions enter through a retarded self-energy $\bm\Sigma^r(\epsilon)$, which modifies the retarded Green’s function according to
\begin{eqnarray}
	\label{eq:G_r}
	\mathbf G^r(\epsilon) = [\epsilon \mathbf I - \mathbf H_{\scriptsize \mbox{s-d}} - \bm \Sigma^r(\epsilon)]^{-1}, 
\end{eqnarray}
where $\mathbf H_{\scriptsize \mbox{s-d}}$ is the matrix representation of the s-d Hamiltonian in the site–spin $(i,\alpha)$ basis. The lesser Green’s function $\mathbf G^{<}$, which is essential for computing physical observables in nonequilibrium steady states, follows from the Keldysh relation:
\begin{eqnarray}
	 \mathbf G^{<}(\epsilon) = \mathbf G^r(\epsilon)\, \bm\Sigma^{<}(\epsilon)\, \mathbf G^a(\epsilon), 
\end{eqnarray}
where $\bm\Sigma^{<}$ is connected to $\bm\Sigma^{r}$ through the fluctuation–dissipation theorem. As an example, the on-site electron density is given by $n_i = \sum_\alpha \langle \hat{c}^\dagger_{i\alpha}\hat{c}^{\,}_{i\alpha} \rangle = \sum_\alpha \int \frac{d\epsilon}{2\pi i} G^{<}_{i\alpha, i\alpha}(\epsilon)$. The exchange field acting on $\mathbf S_i$ in Eq.~(\ref{eq:LL}) is obtained using the generalized Hellmann–Feynman theorem and can be expressed directly in terms of the lesser Green’s function~\cite{stamenova05,salahuddin06,xie17,chern22}:
\begin{eqnarray}
	\label{eq:H_NEGF}
	\mathbf H_i = - \biggl\langle   \frac{\partial \hat{\mathcal{H}}_{\scriptsize \mbox{s-d}}}{\partial \mathbf S_i }  \biggr\rangle 
	= J \sum_{\alpha\beta}\bm\sigma_{\beta\alpha} \int_{-\infty}^{+\infty} \frac{d\epsilon}{2\pi i} \, G^<_{i\alpha, i\beta}(\epsilon). \qquad
\end{eqnarray}
To model the voltage-driven insulator-to-metal transition (IMT) in the s-d system~\cite{chern22}, these NEGF-computed exchange fields are coupled to stochastic LLG dynamics. A small Langevin-type stochastic magnetic field is added at each site to incorporate finite-temperature effects and ensure proper sampling of thermal fluctuations.


The real-space NEGF calculation for even moderately sized lattices (e.g., systems with fewer than $\sim 1000$ spins) is already computationally demanding. The primary bottleneck is the evaluation of the retarded Green’s function $\mathbf G^r(\epsilon)$, which requires inverting a large matrix for each of thousands of energy points $\epsilon$; see Eq.~(\ref{eq:G_r}). In NEGF-LLG simulations of driven itinerant magnets, this entire NEGF procedure must be repeated at every time step of the dynamical evolution, leading to a substantial computational burden. Consequently, only relatively small system sizes can be simulated, despite extensive parallelization efforts. As discussed above, the ML force-field approach provides a powerful strategy for overcoming these limitations. By replacing the repeated, expensive Green’s function calculations with fast ML inference, one can dramatically reduce the cost of multi-scale NEGF-LLG simulations and enable large-scale modeling of nonequilibrium itinerant magnets.

\begin{figure}
\includegraphics[width=1.0\columnwidth]{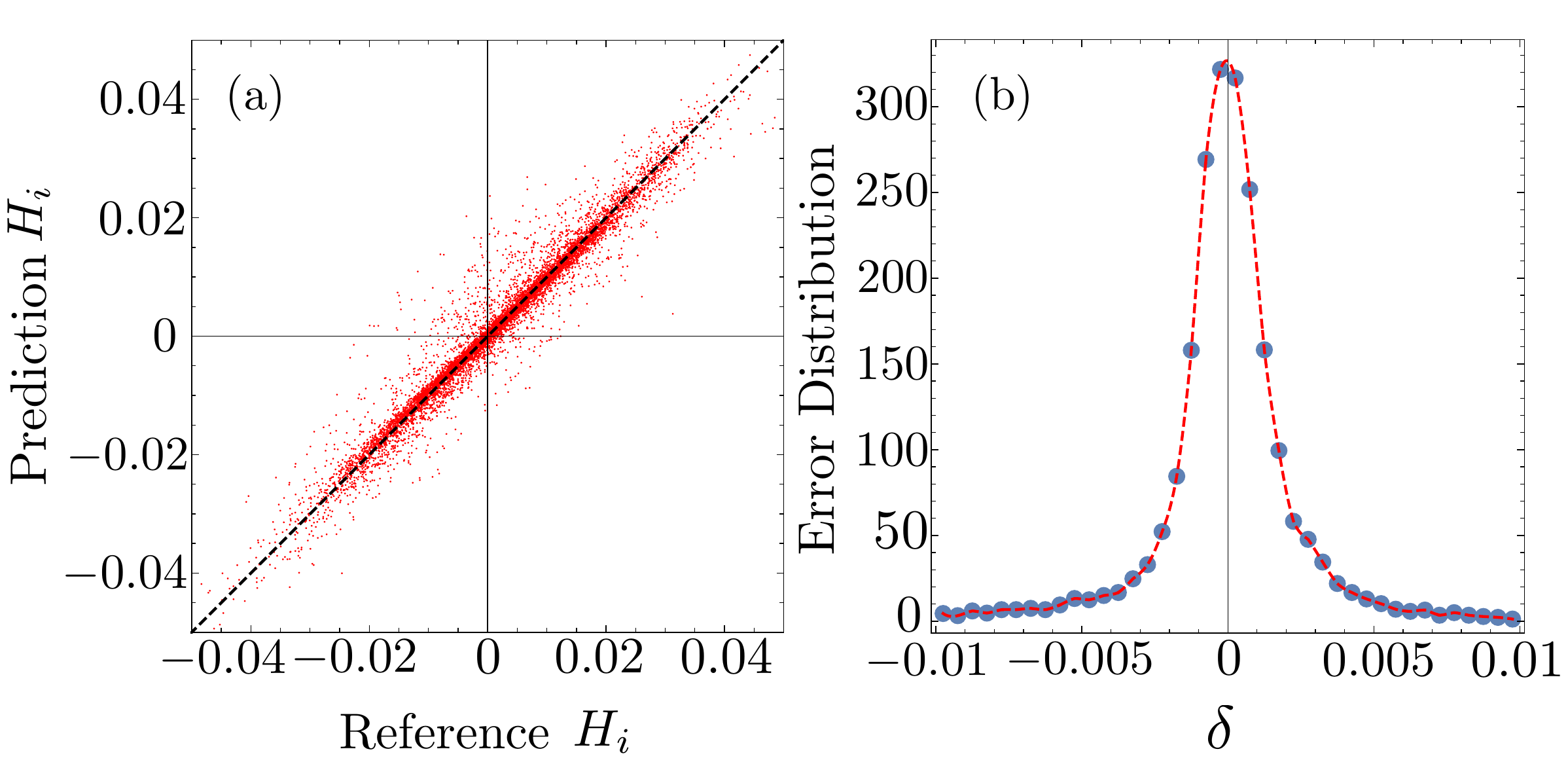}
\caption{(a) The ML predicted forces versus the exact solution from the NEGF calculation for the s-d model with exchange coupling $J = 3.8 t_{\rm nn}$. (b) Normalized distribution of the prediction error of the perpendicular components of the forces.}
\label{fig:negf-forces} 
\end{figure}

To implement the generalized BP framework of Fig.~\ref{fig:generalized-BP}, we construct a six-layer neural network trained directly on the electronic exchange fields obtained from the NEGF calculations~\cite{zhang23}. The training data consist of 3200 snapshots of nonequilibrium electronic states, each providing approximately $600$ force components derived from the generalized exchange-field expression in Eq.~(\ref{eq:force}). In contrast to the standard BP approach—where both forces and total energies contribute to the training loss—we employ a force-only loss function. This choice reflects the fact that a well-defined total energy functional does not exist for open, voltage-driven systems treated within the NEGF formalism. The performance of the trained model is illustrated in Fig.~\ref{fig:negf-forces}(a), which compares the componentwise torques $\mathbf S_i \times \mathbf H_i$ predicted by the neural network with the exact NEGF results. The agreement is excellent, with a mean-squared error of $8.97 \times 10^{-6}$. The inset of Fig.~\ref{fig:negf-forces}(b) shows the normalized distribution of prediction errors obtained from the validation dataset, exhibiting a narrow spread with a standard deviation of $\sigma = 0.0014$.

\begin{figure}
\includegraphics[width=0.99\columnwidth]{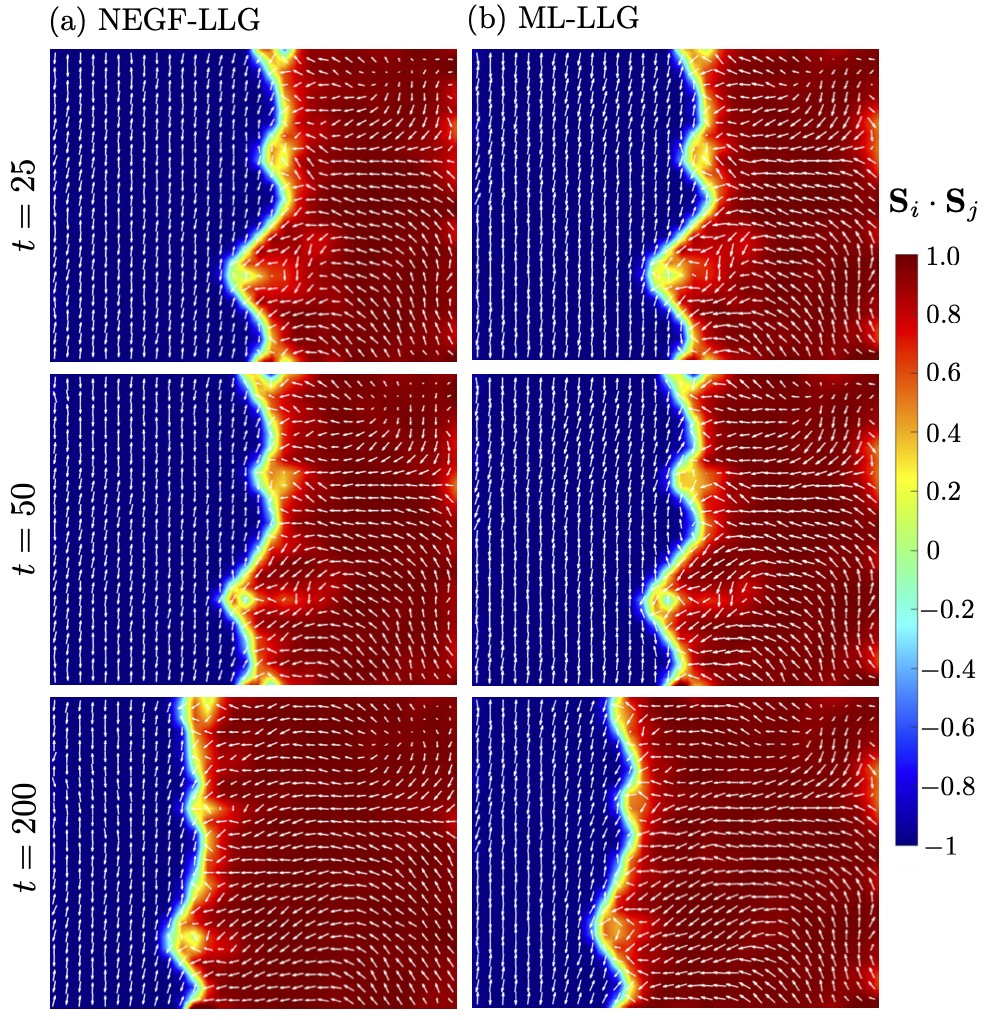}
\caption{Domain wall propagation in a s-d system driven by an external voltage $eV = 3.2 t_{\rm nn}$. Comparison between (a) NEGF-LLG simulations and (b) ML-LLG simulations. The lattice size is $30\times 24$.  The color bar indicates the local nearest-neighbor spin correlation $b_{ij} = \mathbf S_i \cdot \mathbf S_j$. Blue (red) area corresponds to AFM (FM) domains. The NEGF-LLG simulation was carried out at a low temperature of $k_B T = 0.01\,t_{\rm nn}$, while the ML simulation was performed without Langevin noise. 
\label{fig:domain}
}
\end{figure}

We next incorporate the NN-based exchange-field model into the LLG dynamics to simulate voltage-driven domain-wall propagation in the square-lattice s-d system. Figure~\ref{fig:domain} shows snapshots of the bond variable $b_i$, which characterizes nearest-neighbor spin correlations, for both NEGF-LLG and ML-LLG simulations performed on a $30 \times 24$ lattice. Both simulations are initialized from the same configuration containing a well-formed FM-AFM domain wall. A small thermal noise is added to perturb the unstable N'eel background in the driven regime, and the stochastic LLG evolution employs a Langevin thermostat corresponding to a low temperature $k_B T = 0.01 t_{\rm nn}$~\cite{garcia98}.

As discussed above, the kinetics of the nonequilibrium insulator-to-metal transition (IMT) is dominated by the motion of these FM-AFM domain walls. Accordingly, our ML model is trained to accurately predict forces in the interfacial region where the two magnetic phases coexist. In the voltage-driven IMT protocol, the system begins in an insulating antiferromagnetic state with an energy gap $\Delta E_g = 2J$. An external bias $V$ is applied across the two electrodes attached to the left and right boundaries of the sample. When the chemical potential of the right electrode is lowered into resonance with the in-gap edge states, electrons begin to drain from the system into the reservoir, destabilizing the AFM configuration and initiating ferromagnetic ordering at the boundary~\cite{chern22}.
This process nucleates FM domains along the sample’s edge, producing a sharp interface separating two regions of distinct electron density. The insulating AFM background remains half-filled with one electron per site, whereas the emerging FM regions exhibit a significantly reduced local electron density and tend to be metallic. As the applied voltage continues to drive electron extraction, the FM domains expand into the AFM bulk, leading to a progressive growth of the metallic region and ultimately to the voltage-induced transition into a low-resistance state.

\begin{figure}
\includegraphics[width=0.99\columnwidth]{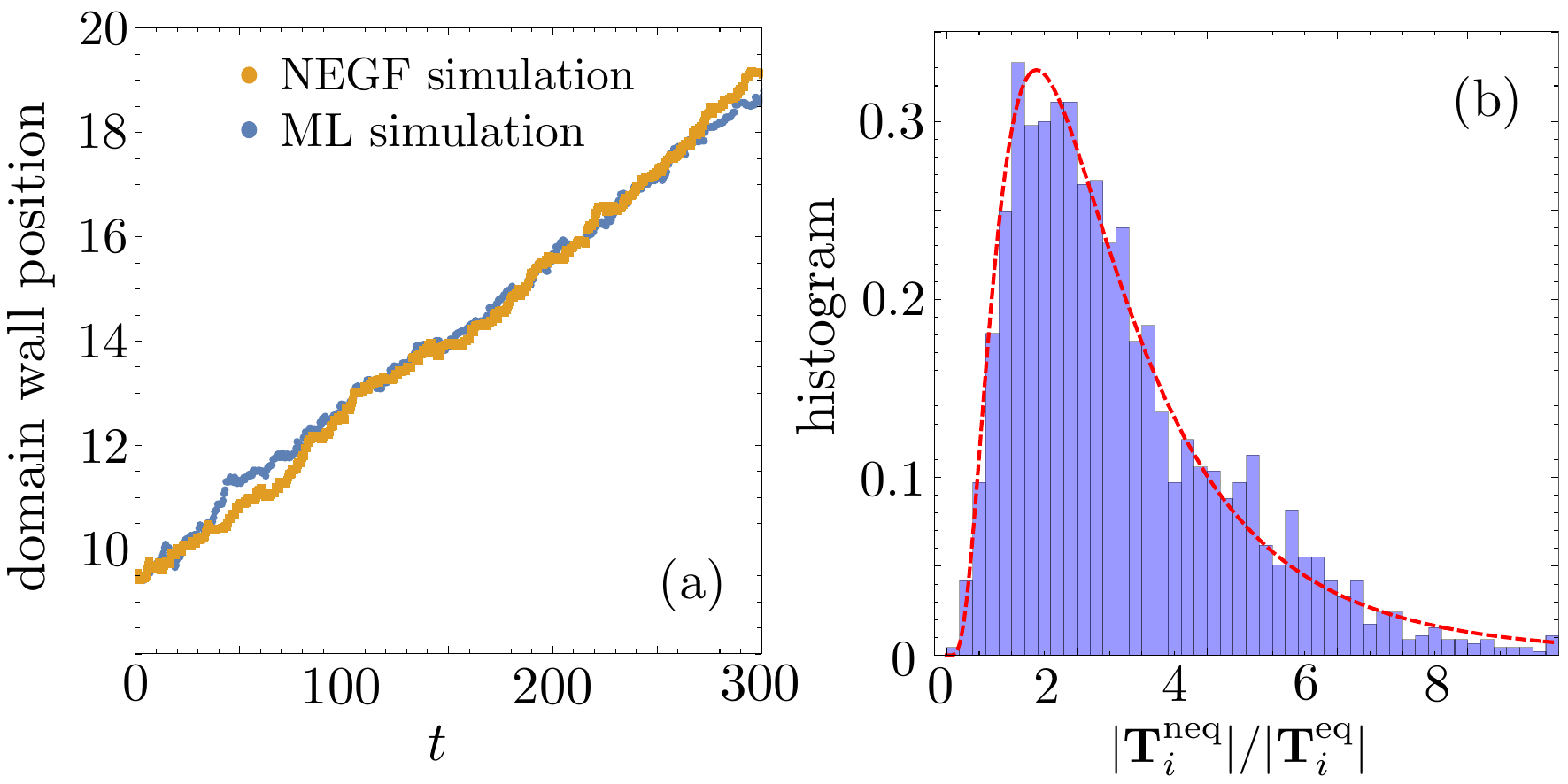}
\caption{(a) Average position of FM-AFM domain, obtained from NEGF-LLG and ML-LLG simulations, as a function of time during the voltage-driven insulator to metal transition of the s-d model.  (b) histogram of the ratio $|\mathbf T^{\rm neq}_i|/ |\mathbf T^{\rm eq}_i|$ predicted by the NN model for the simulation of domain-wall propagation. Here $\mathbf T^{\rm eq}_i = \mathbf S_i \times \partial {E} / \partial \mathbf S_i$ is the quasi-equilibrium torque, and $\mathbf T^{\rm neq}_i =  \mathbf S_i \times (\mathbf S_i \times \partial {G} / \partial \mathbf S_i)$ is the non-equilibrium torque in the generalized LLG Eq.~(\ref{eq:LL2}). 
\label{fig:imt-benchmark} 
}
\end{figure}

Fig.~\ref{fig:imt-benchmark}(a) shows the time evolution of the domain-wall position, averaged over the transverse $y$ direction, for LLG simulations using NEGF-computed forces and ML-predicted forces. The two trajectories exhibit excellent agreement, with only minor deviations arising from the stochastic Langevin noise in the LLG dynamics and the small residual force-prediction error of the ML model. While the close correspondence might suggest that the ML prediction error effectively acts as an additional source of thermal noise, a more plausible interpretation is that thermal fluctuations of this magnitude play only a secondary role. Their primary function is to provide an initial perturbation that destabilizes the N'eel background, whereas the subsequent domain-wall propagation dynamics are largely governed by the underlying nonequilibrium electronic forces.

An important advantage of our NN-based formulation is that it naturally decomposes the electron-mediated exchange field into quasi-equilibrium and nonequilibrium components, $\mathbf h^{\rm eq}$ and $\mathbf h^{\rm neq}$, following Eq.~(\ref{eq:force}). Such a separation is generally not accessible in microscopic calculations. For instance, the NEGF expression in Eq.~(\ref{eq:H_NEGF}) yields only the total exchange field, with no clear way to distinguish conservative from non-conservative contributions. In contrast, introducing the two scalar potentials ${E}$ and ${G}$ in the generalized BP model plays a role analogous to the site energies $\epsilon_i$ in the standard BP framework: although these local quantities cannot be extracted directly from quantum-mechanical calculations such as DFT, the trained ML model can infer them and reveal additional physical structure.
This decomposition enables a direct analysis of the two torque components, $\mathbf T^{\rm eq}_i = \mathbf h^{\rm eq}_i \times \mathbf S_i$ and $\mathbf T^{\rm neq}_i = \mathbf h^{\rm neq}_i \times \mathbf S_i$. Fig.~\ref{fig:imt-benchmark}(b) shows the histogram of the ratio $|\mathbf T^{\rm neq}_i| / |\mathbf T^{\rm eq}_i|$ for spins near the FM–AFM domain interface. As expected, the nonequilibrium torque dominates in this region, confirming that voltage-driven domain-wall propagation is largely controlled by nonequilibrium electronic forces.

The physical roles of the two torque components are illustrated in Fig.~\ref{fig:decomposition}(c). The quasi-equilibrium torque $\mathbf T^{\rm eq}_i = \mathbf S_i \times (\partial {E} / \partial \mathbf S_i)$ drives precessional motion along contours of constant ${E}$. The nonequilibrium torque $\mathbf T^{\rm neq}_i = \mathbf S_i \times [\mathbf S_i \times (\partial {G} / \partial \mathbf S_i)]$ often acts against the conventional Landau–Lifshitz damping torque $\mathbf T^{\rm damping}_i = \lambda, \mathbf S_i \times \mathbf T^{\rm eq}_i$, where $\lambda = \gamma \alpha / (1 + \alpha^2)$ is the effective damping strength. This is supported by the histogram of $(\mathbf T^{\rm neq}_i \cdot \mathbf h^{\rm eq}_i)$ in Fig.~\ref{fig:domain}(e): its predominantly negative values show that the nonequilibrium torque tends to push spins away from the conservative field direction $\mathbf h^{\rm eq}_i = -\partial {E} / \partial \mathbf S_i$, acting in a manner similar to the anti-damping torques discussed in Refs.~\cite{brataas12,ralph08,salahuddin08}.

\section{Conclusion and Outlook}

\label{sec:conclusion}

In summary, we have presented a unified framework for machine-learning (ML) force fields capable of modeling Landau-Lifshitz-Gilbert (LLG) spin dynamics in itinerant electron systems under both quasi-equilibrium and strongly driven conditions. Building on the Behler-Parrinello (BP) paradigm---originally developed for quantum molecular dynamics---we generalized the notion of locality to magnetization dynamics by assuming that the electron-mediated forces acting on each spin depend primarily on its local magnetic environment. This locality principle enables the construction of deep neural networks that accurately encode the highly nonlinear mapping between local spin configurations and effective magnetic fields.

A central ingredient of our approach is the design of symmetry-invariant magnetic descriptors, which provide differentiable and rotationally invariant representations of the local spin environment. We discussed their theoretical underpinnings, emphasized the importance of preserving both lattice point-group and SO(3) spin-rotation symmetries, and presented a practical implementation based on reference irreducible representations. This formulation draws inspiration from group-theoretical power-spectrum and bispectrum methods while remaining computationally efficient.

A key advance of this work is the generalization of the BP architecture to systems driven far from equilibrium. Exploiting the constraint that the magnitude of each spin is conserved in LLG dynamics, we introduced a generalized potential theory that treats both conservative and non-conservative electronic forces on the same footing. This formulation naturally extends BP-type models to nonequilibrium settings and allows ML force fields to represent voltage-driven or current-driven exchange torques in itinerant magnets. As a demonstration, we constructed a neural network that accurately reproduces the electron-mediated forces computed using nonequilibrium Green’s functions (NEGF) for a driven s-d model. The resulting ML-LLG simulations capture the full voltage-induced domain-wall propagation and quantitatively match the NEGF-LLG benchmark results.

Beyond the specific case studied here, the proposed ML framework provides a pathway toward accurate and scalable modeling of spin-transfer torques (STT)~\cite{slonczewski96,berger96,brataas12,ralph08}, which lie at the core of many spintronic technologies. Existing LLG simulations typically employ empirical STT formulas~\cite{slonczewski96,berger96,brataas12,ralph08,bazaliy98,zhang04,tatara08}, but these phenomenological models cannot capture the subtle spin–electron couplings present in real devices. On the other hand, first-principles approaches such as NEGF-based STT calculations are considerably more accurate but remain computationally prohibitive for large spatial domains~\cite{stamenova05,salahuddin06,xie17,ellis17,petrovic18,dolui20,chen09,nikolic20}. ML-based STT models capable of emulating NEGF-level accuracy while retaining linear scaling therefore represent a compelling direction for enabling large-scale dynamical simulations of magnetic textures, device operation, and spintronic architectures.

While we have focused primarily on BP-type constructions, several other ML architectures provide complementary and potentially more flexible routes for modeling itinerant spin systems. Convolutional neural networks (CNNs) naturally reflect local interactions due to their finite receptive fields and have already been applied to scalable spin force-field modeling~\cite{cheng23b,tyberg25}. Recent developments in equivariant neural networks~\cite{thomas2018,anderson2019,townshend2020} embed symmetry directly into the network architecture, while graph neural networks (GNNs) achieve linear scaling through message passing and symmetry-preserving graph transformations~\cite{scarselli09,reiser2022,batzner2022,batatia2025}. Extending these architectures to model LLG dynamics in itinerant magnets is a particularly promising direction for future exploration.

Finally, the advent of fully SO(3)-equivariant neural networks~\cite{weiler18,batzner22,gong23} offers a powerful alternative to descriptor-based models. Recent work demonstrates that spin-rotation symmetry can be enforced directly within an equivariant convolutional architecture~\cite{Miyazaki23}. However, an outstanding challenge remains: how to treat both continuous spin-rotation symmetry and discrete lattice point-group symmetry in a unified equivariant framework. Addressing this question represents an important frontier for developing next-generation ML force fields for magnetic materials.

\begin{acknowledgments}
This work was supported by the US Department of Energy Basic Energy Sciences under Contract No. DE-SC0020330. The authors acknowledge Research Computing at The University of Virginia for providing computational resources and technical support that have contributed to the results reported within this publication. 
\end{acknowledgments}

\bibliography{ref.bib}

\end{document}